\documentclass[lettersize,journal]{IEEEtran}
\usepackage{amsmath,amsfonts}
\usepackage{algorithmic}
\usepackage{algorithm}
\usepackage{array}
\usepackage[caption=false,font=normalsize,labelfont=sf,textfont=sf]{subfig}
\usepackage{textcomp}
\usepackage{stfloats}
\usepackage{url}
\usepackage{verbatim}
\usepackage{graphicx}
\usepackage{cite}

\usepackage{amsmath,amsfonts}
\usepackage{algorithmic}
\usepackage{graphicx}
\usepackage{textcomp}
\usepackage{xcolor}
\usepackage{soul}
\usepackage{booktabs}
\usepackage{rotating}
\usepackage{multirow}
\usepackage{diagbox}
\usepackage{array}
\usepackage{threeparttable}
\usepackage{booktabs}
\usepackage[framemethod=tikz]{mdframed}
\usepackage{threeparttable}
\usepackage{xcolor}
\usepackage{tikz}
\usepackage[T1]{fontenc}

\newcommand{\circled}[1]{%
  \begin{tikzpicture}[baseline=(char.base)]
    \node[shape=circle, draw, inner sep=1pt, minimum size=0.6em] (char) {\small #1};
  \end{tikzpicture}%
}

\mdfdefinestyle{graystyle}{
    backgroundcolor=gray!20, 
    linecolor=gray!20, 
    linewidth=0pt 
}

\begin{document}

\title{SoK: Systematizing Software Artifacts Traceability via Associations, Techniques, and Applications}

\author{Zhifei Chen, Lata Yi, Liming Nie, Yangyang Zhao, Hao Liu, Yiqing Shi, and Wei Song,~\IEEEmembership{Senior Member,~IEEE}

\thanks{Zhifei Chen, Lata Yi, Yiqing Shi, and Wei Song are with the School of Computer Science and Engineering, Nanjing University of Science and Technology, China (e-mail: chenzhifei@njust.edu.cn; yilata1241@njust.edu.cn; 125106010860@njust.edu.cn; wsong@njust.edu.cn).}
\thanks{Liming Nie and Hao Liu are with the School of Artificial Intelligence, Shenzhen Technology University, China (e-mail: nieliming@sztu.edu.cn; 2410263026@stumail.sztu.edu.cn).}
\thanks{Yangyang Zhao is with the School of Computer Science and Technology, Zhejiang Sci-Tech University, China (e-mail: yangyangzhao@zstu.edu.cn).}
\thanks{Lata Yi contributed equally as the co-first author.}
\thanks{Liming Nie is the corresponding author.}}

\markboth{Journal of \LaTeX\ Class Files,~Vol.~14, No.~8, August~2021}%
{Shell \MakeLowercase{\textit{et al.}}: A Sample Article Using IEEEtran.cls for IEEE Journals}


\maketitle

\begin{abstract}
Software development relies heavily on traceability links between various software artifacts to ensure quality and facilitate maintenance. While automated traceability recovery techniques have advanced for different artifact pairs, the field remains fragmented with an incomplete overview of artifact associations, ambiguous linking techniques, and fragmented knowledge of application scenarios. To bridge these gaps, we conducted a systematic literature review on software traceability recovery to synthesize the linked artifacts, recovery tools, and usage scenarios across the traceability ecosystem. First, we constructed the first global artifacts traceability graph of 23 associations among 22 artifact types, exposing a severe research imbalance that heavily favors code-related links. Second, while recovery techniques are shifting toward deep semantic models, a reproducibility crisis persists (e.g., only 37\% of studies released code); to address this, we provided a comprehensive evaluation framework including a technical decision map and standardized benchmarks. Finally, we quantified an industrial adoption gap (i.e., 95\% of tools remain confined to academia) and proposed a role-centric framework to dynamically align artifact paths with concrete engineering activities. This review contributes a coherent knowledge framework for artifacts traceability research, identifies current trends, and provides directions for future work.
\end{abstract}


\section{Introduction}
Software development is an inherently complex process that generates a multitude of artifacts, ranging from initial requirement and design models to code, tests, and deployment architectures \cite{52,17,29}. To satisfy the increasing regulation for safety- or security-critical systems, establishing and maintaining clear relationships, or ``traceability links'', among diverse artifacts is paramount. For example, robust traceability provides the necessary evidence to demonstrate that software design fulfills all specified software requirement \cite{19}, and all code is linked to well-defined specifications and established testing procedures \cite{15,17,33}. Furthermore, it supports a variety of software tasks, such as change impact analysis \cite{aung2020literature}, bug-fix commit identification \cite{nguyen2022hermes}, selective regression testing \cite{naslavsky2007using}, and project management \cite{panis2010successful}. Conversely, the absence of these links forces engineering teams to comprehend, validate, and evolve complex systems blindly. Thus, traceability is not merely an optional benefit, but a fundamental engineering necessity to prevent structural decay during software evolution.

However, due to the dynamic nature of software projects, manual creation and maintenance of links between software artifacts are often labor-intensive, error-prone, and frequently neglected, leading to ``traceability debt'' \cite{rodriguez2021leveraging}. Consequently, automated or semi-automated recovery of software artifacts traceability has emerged as a critical research area in recent years. 
Researchers have explored various techniques, including information retrieval (IR) methods \cite{wang2022systematic}, machine learning (ML) approaches \cite{wang2023systematic}, and heuristic-based algorithms \cite{11,18,23}, to identify and reconstruct links between disparate software artifacts. Existing studies often focus on a particular pair of artifacts, such as ``\textit{requirement} - \textit{code}'' \cite{53,54,55,56,57} and ``\textit{test} - \textit{code}'' \cite{34,35,36}. Although these advances have shown promising results in specific contexts, the field remains fragmented with diverse artifact relationships, recovery techniques, evaluations, and scopes. This leads to a lack of a holistic overview of the practical applicability of these recovered links in real-world development scenarios. 

Following these previous studies, we can find that tackling the topic of software traceability recovery presents vital challenges in conceptual, technical, and practical aspects:

\textbf{\textit{(1) Incomplete Overview of Artifact Associations:}} Current research often focuses on specific pairs of artifacts or limited sets of relationships. This incomplete landscape leaves researchers operating with severe blind spots. Without knowing which critical associations are missing, the community cannot strategically prioritize new artifact pairs to improve tool coverage or support more complex engineering scenarios.

\textbf{\textit{(2) Ambiguous Techniques and Evaluations:}} Despite various tools being developed, the underlying linking techniques and evaluation boundaries are reported inconsistently across studies. This methodological ambiguity prevents fair and cross-study comparisons, making it nearly impossible to establish clear state-of-the-art baselines or for practitioners to select the optimal technique for their specific context.

\textbf{\textit{(3) Fragmented Knowledge of Practical Applications:}} While theoretical benefits of recovered links are widely acknowledged, their real-world application remains unclear. This lack of synthesized usage scenarios creates a significant barrier to industrial adoption. Engineering teams are unclear in which specific lifecycle phases or tasks these links will deliver concrete practical benefits or alleviate actual maintenance efforts.

Motivated by the urgent need to resolve these bottlenecks, this paper conducts a systematic literature review to bridge these critical gaps. We aim to answer three questions: \textit{RQ1) What types of software artifacts and their associations constitute the current traceability networks? RQ2) What is the current status of existing tools to establish links between software artifacts? RQ3) What are the usage scenarios of the recovered links between software artifacts?}

To investigate these questions, we adopted the research methodology of Systematization of Knowledge (SoK). In particular, we conducted a comprehensive search for relevant literature published in major academic conferences and journals. Through rigorous selection, classification, and synthesis of 76 selected literatures, we carefully identified associations between artifacts, core linking techniques, and usage scenarios. This study uncovers several critical empirical findings. We identified 22 distinct software artifacts and 23 types of associations, but the research landscape is highly unbalanced: nearly half of the studies focus on the consistency between \textit{documentation} and \textit{source code}. Technically, the field exhibits a clear paradigm shift from traditional IR methods to advanced learning models to achieve superior semantic comprehension. Despite this advancement, reproducibility remains a major bottleneck: only 37\% of studies made their source code publicly available and researchers still lack standardized evaluation benchmarks. Furthermore, 95\% of studies were evaluated in academic settings, primarily targeting requirement - implementation consistency and maintenance support, which exposes a massive gap in real-world industrial adoption.

Beyond simply reporting these empirical findings, this paper provides actionable frameworks and strategic insights for the community. Our primary contributions are manifold:

\textbf{\textit{(1) Artifacts Traceability Graph:}} We constructed the first global artifacts traceability graph that synthesizes 23 distinct relational associations among 22 hierarchical software artifacts. Building on this holistic map, we formulated structural guidelines (including multi-hop chains, the central pivot, and quality boundaries) to shift the research focus from isolated binary mappings toward a complete ecosystem perspective.

\textbf{\textit{(2) Technical Landscape and Evaluation Frameworks:}} We revealed the technical paradigm shift from traditional IR to deep semantic models. To address the reproducibility crisis in this field, we proposed a technical decision map and a comprehensive evaluation framework to standardize benchmarks.

\textbf{\textit{(3) Goal-Driven Traceability Framework:}} By mapping the application domains of recovered links, we provided quantitative evidence of the disconnect between academic research and real-world industrial utility. To bridge this gap, we introduced a role-centric traceability framework that dynamically aligns specific artifact paths with concrete engineering objectives.

All materials related to this paper are publicly available \cite{questionnaire}.
The remainder of this paper is organized as follows. Section \ref{background} presents the background and related work. Section \ref{design} details our review methodology, and Section \ref{results} presents the corresponding results. Section \ref{discussion} discusses the findings and limitations of this study. Finally, Section \ref{conclusion} concludes this paper and outlines future work.

\section{Background and Related Work}\label{background}
\subsection{Software Traceability Recovery}
Software artifacts are all tangible byproducts generated throughout the software lifecycle. To our knowledge, there is no comprehensive list of software artifacts defined to date. These diverse outputs encompass a wide range of deliverables, all essential for defining and supporting a software system.
Software artifacts traceability recovery is the process of automatically identifying and establishing the relationships between these disparate software artifacts.
In the complex landscape of modern software development, it is necessary to maintain a clear and comprehensive understanding of the relationships between various artifacts \cite{charalampidou2021empirical}. Software traceability ensures that every component of a system can be linked back to its origin and forward to its impact, providing a crucial foundation for effective project management \cite{wan2023software}.

However, manually establishing and maintaining artifact links throughout the software lifecycle is labor-intensive and error-prone. As projects grow in size and complexity and as development teams become more distributed, the sheer volume of artifacts and the dynamic nature of their interdependencies make manual traceability increasingly impractical \cite{castellanos2022compliance}. This challenge gives rise to the critical need for the research of artifacts traceability recovery. 


\subsection{Related Work}

\subsubsection{Establishment of Artifacts Traceability Links}
Numerous studies have explored methods for establishing traceability links between different software artifacts.
In the initial phase, traceability was maintained manually by qualified developers who were responsible for creating and updating trace links between software artifacts. For example, Alves-Foss et al. \cite{994466} established trace links between UML design specifications and corresponding source code using XML technology, supporting hypertext-based traceability through manual means.
Afterwards, semi-automated methods were proposed to reduce manual workload while retaining human oversight for critical decisions. For example, Hammad et al. \cite{hammad2011automatically} proposed an automated technique to determine whether changes in source code affect UML class diagrams in design documents. The system notified users when specific decisions were required.

In recent years, researchers have been conducting studies towards fully automated traceability techniques. 
Abadi et al. \cite{abadi2008traceability} compared five traditional IR techniques (LSI, VSM, JSM, PLSI, and SDR) for traceability between \textit{code} and \textit{documentation}. They concluded that VSM and LSI are the most suitable for traceability recovery tasks, despite their relatively poor performance in dimensionality reduction.
Cleland-Huang et al. \cite{1} employed a classifier model trained on manually curated traceability matrices to establish links between \textit{regulation} and \textit{requirement}. Rahimi et al. \cite{59} addressed the issue of low accuracy in automated traceability between software \textit{requirement} and \textit{source code} by leveraging a neural network-based semantic vectorization approach. Chen et al. \cite{47} improved the accuracy and stability of ``\textit{documentation} - \textit{code}'' traceability by integrating machine learning, and heuristic optimization techniques. Although numerous techniques exist for associating software artifacts, there is a clear need for a comprehensive synthesis of these approaches.

\subsubsection{Existing Reviews of Artifacts Traceability Recovery}
Beyond individual studies on different artifacts linking approaches, several reviews have systematically assessed the broader landscape of artifacts traceability establishment.


Several reviews concentrated on a specific artifact to highlight its unique role across the software lifecycle \cite{cybulski1998reuse}. For example, Parizi et al. \cite{parizi2014achievements} presented a systematic literature review focusing on tests, addressing a knowledge gap in the recovery of traceability from test to code. Similarly, Wang et al. \cite{wang2024advancements} explored trends and advances in bug traceability, concluding that improving the accuracy of ``\textit{bug} - \textit{commit}'' and ``\textit{bug} - \textit{code}'' linking is essential for software systems.

Other reviews examined traceability across multiple software artifacts within the scope of a specific research domain or a specific technique.
For example, Aung et al. \cite{aung2020literature} conducted a systematic review on automated trace link recovery methods within the domain of change impact analysis, concluding that there is a lack of public datasets and that traceability research remains limited in these areas. 
Wang et al. \cite{wang2023systematic} focused on the intersection of ML and software traceability, providing a comprehensive survey that considered the classification of artifact links as a key research area. 
Charalampidou et al. \cite{charalampidou2021empirical} conducted a study of previous artifact-related surveys and concluded that requirement artifacts dominate the traceability literature with most surveys.
Recent systematic reviews highlight a decisive paradigm shift toward deep learning. Khalil et al. \cite{khalil2025systematic} and Rahman et al. \cite{abd2025exploring} noted that advanced embedding and Transformer-based architectures (e.g., BERT) have become the dominant approach for capturing complex semantic relationships, appearing in over half of the latest traceability studies. Complementing this, the comprehensive review by Antonio et al. \cite{rosado2025machine} on machine learning techniques for requirement engineering points out that although supervised learning is widely applied to tasks such as requirement classification and traceability across the requirement engineering lifecycle, its effectiveness is often constrained by the scarcity of high-quality labeled datasets. Furthermore, Koboyatshwene et al. \cite{koboyatshwene2025requirements} identified persistent gaps in the literature, particularly the ongoing neglect of non-functional requirement and the lack of standardized benchmarks for global artifact sets.
However, these existing reviews focus only on limited artifacts, single domains, or single techniques, without providing a systematic overview of this research area.
These limitations underscore the need for a more general framework to bridge the semantic gap across diverse and global software artifacts.

\section{Research Design}\label{design}
In this section, we outline the research questions and then present the detailed description of our research framework.
\subsection{Research Questions}
Our study aims to answer the following research questions.

\textbf{RQ1: What types of software artifacts and their associations constitute the current traceability networks?} 
Rather than viewing traceability as isolated point-to-point mappings, this question explores the multidimensional associations between heterogeneous artifacts. We aim to construct a global traceability network, pinpointing established research focuses and highlighting structural blind spots within the ecosystem.

\textbf{RQ2: What is the current status of existing tools to establish links between software artifacts?} 
Recognizing the structural complexity of the traceability network (RQ1), this question investigates the technical foundations required to construct it. By comprehensively analyzing artifact representations, linking techniques, and evaluation frameworks, we aim to understand how existing tools bridge heterogeneous semantic gaps and to assess the maturity of current empirical evaluations, providing guidance for future tool selection.

\textbf{RQ3: What are the usage scenarios of the recovered links between software artifacts?} 
Even when successfully constructed using advanced techniques (RQ2), the traceability network is prone to decay if it lacks clear practical purposes. Therefore, this question seeks to understand how the recovered links align with specific domains, software phases, and practical objectives. Identifying these scenarios provides crucial insights to bridge the gap between academic research and real-world industrial adoption.

\subsection{Research Framework}
\begin{figure}
    \centering
    \includegraphics[width=1\linewidth]{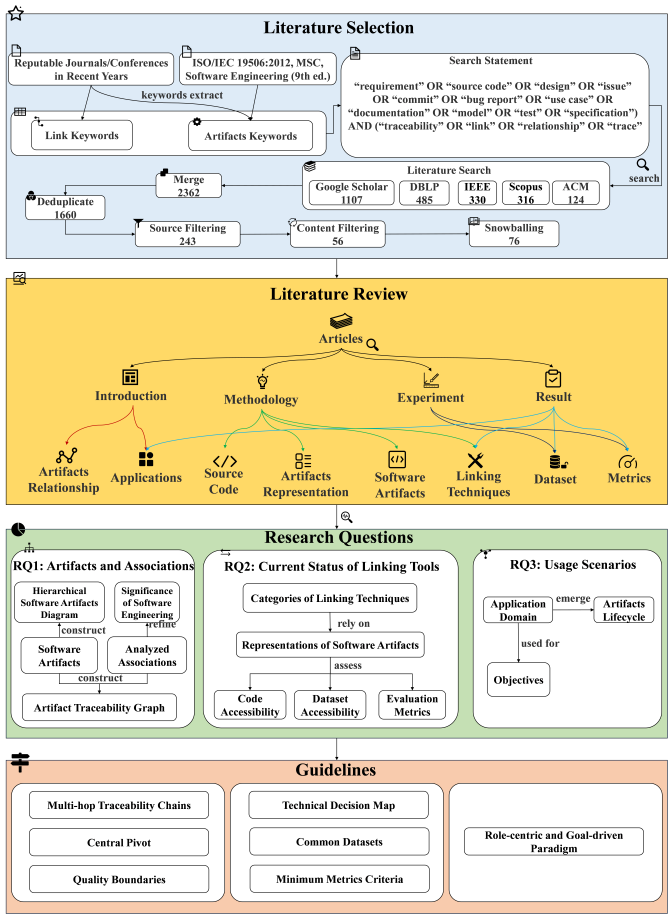}
    \caption{Overview of the Study Framework.}
    \label{fig:framework}
\end{figure}

To address these research questions, our study follows the general guidelines for the preceding systematic reviews proposed by Kitchenham and Barbara \cite{kitchenham2004procedures}. The general flow of the process is illustrated in Fig. \ref{fig:framework}. 
First, the Literature Selection module selected a list of relevant papers on software traceability recovery from various sources in recent years.
We believe that these papers can reflect the research trends and encompass the majority of relevant associations between different artifacts. Second, the Literature Review module extracted information from each selected paper. Finally, we summarized the associations between artifacts analyzed in the literature to construct an artifacts traceability graph to answer RQ1. Meanwhile, we analyzed the linking tools and the applications in current research to answer RQ2 and RQ3, respectively. We describe the methodology in the following subsections. More details are available publicly \cite{questionnaire}.

\subsection{Systematic Literature Review}
\subsubsection{Search Query Generation}
In our review protocol, we intend to utilize reputable literature search engines and databases to identify high-quality research papers. Considering the scope of our literature review, we concentrated on a specific set of keywords to perform the paper search. Our search query is structured as a conjunction of two research domains: \textit{D1}) \textit{Software Artifacts} and \textit{D2}) \textit{Link}. Each domain within the search string is expressed as a disjunction of its associated keywords. Our search query \textit{Q} is defined as follows:
$$
\mathit{Q} = \bigwedge_{d\in\left\{D_{1},D_{2}\right\}}\left(\bigvee_{\mathit{keyword }\in K_{d}}\mathit{keyword }\right)
$$
where $K_{d}$ is the set of keywords for the domain \textit{d}.

However, these keywords appear in the literature in different forms. In particular, the scope of \textit{D1} is large and \textit{D2} contains multiple synonymous words. Therefore, to ensure a comprehensive search, we first manually summarized all keywords by checking high-quality research papers in recent years. We collected all research papers from five journals and conferences (\textit{RE}, \textit{ICSME}, \textit{ICPC}, \textit{ICST}, and \textit{SST}) related to our research topic
published since 2022. Subsequently, for each paper, we carefully analyzed whether it investigated the establishment of traceability links between different types of software artifacts. This analysis was conducted with reference to existing formal descriptions and classification standards of software artifacts, such as \textit{ISO/IEC 19506:2012} \cite{ISOIEC}, \textit{Model Services Contract (MSC)} \cite{MSC}, and \textit{Software Engineering (9th ed.)} \cite{SoftwareEngineering}. During this step, we identified 18 relevant papers in the scope of this study. After that, we manually extracted the domain keywords from their titles, abstracts, and author keywords, and finally collected 13 individual keywords for both \textit{D1} and \textit{D2}. Meanwhile, we also included all common software artifacts listed in the formal definitions \cite{MSC,SoftwareEngineering,ISOIEC} into the set of keywords for \textit{D1}.
After combining four keyword sources and subsequently eliminating duplicate and subsumed keywords, we were able to identify 11 and 4 specific keywords for \textit{D1} and \textit{D2}, respectively.
This preliminary analysis provided basic knowledge for expressing the search query string \textit{Q}:
\smallskip
\hrule
\smallskip
\noindent(\textit{``requirement''} \textbf{OR} \textit{``source code''} \textbf{OR} \textit{``design''} \textbf{OR} \textit{``issue''} \textbf{OR} \textit{``commit''} \textbf{OR} \textit{``bug report''} \textbf{OR} \textit{``use case''} \textbf{OR} \textit{``documentation''} \textbf{OR} \textit{``model''} \textbf{OR} \textit{``test''} \textbf{OR} \textit{``specification''})  \textbf{\quad AND \quad}  (\textit{``traceability''}  \textbf{OR} \textit{``link''} \textbf{OR} \textit{``relationship''} \textbf{OR} \textit{``trace''})
\smallskip
\hrule
\smallskip
\subsubsection{Literature Search}
Using the carefully generated \textit{Q}, we conducted a literature search in five authoritative databases including Google Scholar, DBLP, IEEE, Scopus, and ACM. Our search was restricted to titles, abstracts, and author keywords. 
After this systematic search, the number of candidate literature retrieved is: Google Scholar (1107), DBLP (485), IEEE (330), Scopus (316), and ACM (124). This search process was completed by November 7, 2025.
To maintain the integrity of our review, we merged them and removed duplicate entries from the initial pool of literature. It led to a set of 1,660 papers to be evaluated in the next stage.
\subsubsection{Literature Filtering}
To ensure the quality, quantity, and relevance of the selected literature to our research topic, we conducted a two-stage filtering process for the collected papers: source filtering and content filtering.

\textbf{Source Filtering.}
The purpose of paper source filtering was to improve the quality and relevance of collected papers. Our source selection was based on a dual criterion:

\textit{(1)} Specialized journals/conferences highly related to software traceability recovery: \textit{RE}, \textit{ICSME}, \textit{ICPC}, \textit{ICST}, and \textit{SST}.

\textit{(2)} Top-tier core journals/conferences in software engineering: \textit{TSE}, \textit{TOSEM}, \textit{EMSE}, \textit{JSS}, \textit{IST}, \textit{SPE}, \textit{IEEE Software},
\textit{ICSE}, \textit{FSE}, \textit{ASE}.

After combining these criteria, the final set of sources from which we selected papers included 15 reputable journals/conferences. Through this dual strategy, we captured the most authoritative research (consisting of 243 papers) while ensuring deep coverage of the specialized area of traceability.

\textbf{Content Filtering.}
The purpose of paper content filtering was to improve the relevance of the research content. We applied the following inclusion criteria to select papers:

\textit{(1)} Papers whose topic is software traceability recovery.

\textit{(2)} Peer-reviewed journal/conference papers.

\textit{(3)} Papers written in English.

\textit{(4)} Papers with full text available.

\textit{(5)} Papers that explicitly describe linking techniques.

\textit{(6)} Papers that establish links between software artifacts.

To ensure the rigor of the study, two authors independently conducted content filtering. They first jointly excluded studies that were clearly not related to the research topic. The remaining papers were then reviewed in full to determine whether they addressed software traceability recovery. For the two papers that had conflicting decisions, a third author participated in the discussion and made the final judgment. After this filtering process, we obtained 56 relevant papers.

\subsubsection{Snowballing}
To include more essential literature, we employed a bidirectional snowball strategy to trace both backward and forward references of the 56 relevant papers. 
In particular, we considered papers that were cited by these papers or that cited them, extending our search to a two-layer citation depth to maintain focus. Throughout the entire snowballing process, we consistently applied the source filtering and content filtering described in the previous subsection.
This process identified 20 additional relevant papers, bringing 76 papers for the follow-up step.

\subsection{Artifacts Traceability Graph Construction}
To support a systematic overview of different artifacts associations, we built an artifacts traceability graph to visualize the associations that have been a research focus.

\textbf{Software Artifacts (Nodes) Construction.}
The construction of artifact nodes involved three steps: extracting analyzed artifacts from the reviewed literature, removing redundant
artifacts, and constructing an artifact hierarchy. The whole process was manually verified by at least two authors.

First, we carefully reviewed the ``Methodology'' section of each paper collected, and identified the analyzed software artifacts according to the formal definition \cite{ISOIEC,MSC}. This process resulted in a total of 152 software artifacts.

To eliminate redundancy, we simplified the list of artifacts by consolidating software artifacts. We merged coarser-grained terms that referred to the same artifact (e.g., test and test artifact) across the software lifecycle. 
This step results in the identification of 20 distinct software artifacts.

Following that, we proceeded to identify the hierarchy of software artifacts. The artifacts referenced in the guidelines \cite{MSC} served as the foundation for the artifact hierarchy, providing the primary structure. 
First, we extracted artifact categories from the guidelines based on their granularity to form the top layer of the hierarchy. Then, each identified artifact was carefully analyzed in terms of its function within the software development lifecycle. Based on both its granularity and functional role, each artifact was assigned to the most appropriate position within the remaining hierarchical layers.

\textbf{Traceability Links (Edges) Construction.}
The nature of the relationships between different artifacts largely determines whether such links can be meaningfully established. Building upon the generated software artifact hierarchy, we systematically examined the ``Introduction'' sections of selected papers to extract descriptions concerning the targeted connections between software artifacts. For the papers that focused on the same pair of software artifacts, we compared their descriptions of analyzed traceability links to infer whether they refer to the same relationships between artifacts. Finally, distinct link types can be identified after synthesizing and categorizing the underlying relationships targeted in all papers. 

\textbf{Graph Generation.}
After completing the construction of software artifacts and traceability links, we can build the artifacts traceability graph, where artifacts are represented as nodes and the relationships between them as edges. This graph systematically illustrates the interconnections among different types of software artifacts that have been a focus of research. Note that the artifact nodes are arranged hierarchically within their respective software artifact groups in the graph.

\textbf{Expert Survey.}
To validate the constructed artifacts traceability graph and the related findings, we conducted an expert evaluation involving software engineers and researchers using two complementary methods: a structured online survey and in-depth expert interviews. The online survey evaluated the rationality, completeness, and practical value of our research.
Participants in this survey were required to accept a 30-minute training based on our material provided. The questionnaire consisted of 12 questions, where seven questions used a Likert scale from 1 (low) to 5 (high) \cite{albaum1997likert} to assess different dimensions of the graph and our findings, four questions collected demographic information from participants, and one question verified whether they had carefully read the provided materials.
The survey was carried out on the Wjx.cn platform from Jan. 21, 2026 to Feb. 20, 2026. We invited developers of open-source projects and authors of academic papers to participate in it. A total of 38 responses have been received. The complete questionnaire is available on our website \cite{questionnaire}.

Table \ref{tab:demographic} presents the demographic information of all online survey participants, including age, years of professional experience, occupational background, and familiarity with the concept of software traceability recovery. The results show that 76\% of participants have more than six years of professional experience. 92\% of participants rated their understanding of software traceability recovery as 4 or 5, indicating a high level of expertise. Their professional background provides a solid foundation for evaluating the effectiveness of our research.

The expert interviews aimed to evaluate the value of our research in real-world software development and research contexts. Open-ended questions were designed to gather expert insights on our research and how it could help optimize software development or research processes. The specific question posed was: ``\textit{Based on the Artifacts Traceability Graph and the Hierarchical Software Artifacts Diagram, what is the specific value of our findings in tackling the primary issues in your professional field or research area?}'' The interviews were conducted between Feb. 15, 2026 to Feb. 30, 2026 and involved four experts. All data were anonymized and privacy protection measures were implemented to ensure confidentiality. 


\begin{table}[]
\centering
\caption{Demographic Information of Survey Participants.}
\label{tab:demographic}
\resizebox{0.48\textwidth}{!}{
\begin{tabular}{cccc}
\hline
\textbf{Information} & \textbf{Measure} & \textbf{Number} & \textbf{Percentage} \\ \hline
\multirow{5}{*}{Age}& 18-25& 1& 2.63\%\\
 & 26-30& 20& 52.63\%\\
 & 31-40& 7& 18.42\%\\
 & 41-50& 7& 18.42\%\\
 & 51-60& 3& 7.89\%\\ \hline
\multirow{5}{*}{Years of professional experience} & 0-2& 0& -\\
 & 3-5& 9& 23.68\%\\
 & 6-10& 20& 52.63\%\\
 & 11-15& 4& 10.53\%\\
 & Over 16 & 5 & 13.16\% \\ \hline
\multirow{3}{*}{Occupation} & Academic researcher & 14 & 36.84\% \\
 & Software engineer & 21 & 55.26\% \\
 & Oher & 3 & 7.89\% \\ \hline
\multirow{5}{*}{Familiarity with this topic}  & 5(Familiar) & 12 & 31.58\% \\
 & 4 & 18 & 47.37\% \\
 & 3 & 8 & 21.05\% \\
 & 2 & 0 & - \\
 & 1(Unfamiliar) & 0 & - \\ \hline
\end{tabular}
}
\end{table}


\subsection{Analysis of Linking Tools}
We analyzed the tools used to establish links between software artifacts in selected papers. To streamline this process and ensure the focus of our analysis, we defined the following inclusion criteria.

\textit{(1) Applicability:} The tool must be explicitly introduced to establish links between software artifacts, using terms such as ``identify'', ``establish'', or ``link''.

\textit{(2) Automation:} The tool must offer automated software traceability recovery capabilities.

\textit{(3) Description:} The tool must be supported by clear and comprehensive documentation.

For each included tool, we performed a comprehensive analysis in three key dimensions: linking technique, input representation, and evaluation experiments. 
Our tool analysis relies exclusively on the datasets, descriptions, and results reported in the selected literature, rather than empirical execution or independent benchmarking.
First, we examined ``Methodology'' sections of relevant papers to identify the main technical method employed by each tool. We extracted details on the linking algorithm or workflow, giving priority to the choice of the proposed linking technique that achieved the highest performance as reported in ``Results'' sections. 
Second, we identified the types of artifact input representations based on the information provided in ``Methodology'' sections. Certain representation types are suitable only for specific kinds of artifacts; e.g., Abstract Syntax Trees are often used to represent source code. The papers we reviewed exhibit a wide range of artifact representations, also depending on the linking techniques employed. During our examination of each tool, we recorded the representation format used for each software artifact whenever a link was established. The input formats of two artifacts determine how to compute their similarity to support traceability recovery.
Third, in order to assess the reproducibility of the study, we examine the ``Methodology'' section of each paper to identify any references to executable source code.
We also analyzed the evaluation design for each tool, focusing on the datasets and evaluation metrics reported in the sections ``Experiment'' or ``Results''.

\subsection{Analysis of Usage Scenarios}
To understand the usage scenarios in which links between software artifacts are established, we investigated the specific application context of building artifacts links for each paper.
Given that software artifacts emerge at different stages of software lifecycle and serve varying roles, we investigated three dimensions: research domain, software lifecycle phase, and primary objective.

We collected usage scenario information from ``Introduction'' and ``Result'' sections where contributions are typically summarized. We also refer to the discussions of applications described in the papers.
First, we checked whether the software they analyzed contain general projects across various domains or contain industrial projects in specific domains. Through that, we can determine whether the proposed method was applied in an academic or industrial setting. Second, we inferred the phase of software lifecycle emphasized in the study, based on the nature of associated artifacts. Lastly, we extracted and summarized the main objective of recovering traceability links from each study.
This three-dimensional analysis provides an overview of the application contexts in which different pairs of software artifacts traceability should be explored.

\section{Research Results}\label{results}
\subsection{(RQ1) Artifacts and Associations}
After linking and grouping the identified artifacts from relevant papers, we built hierarchical artifacts diagram shown in Fig. \ref{fig:tree} and the traceability graph shown in Fig. \ref{fig:traceability graph}. In the traceability graph, the background colors represent different groups of artifacts, and the node colors represent different hierarchical layers of artifacts in each group. In the following, we describe the findings in these figures.

\subsubsection{Investigated Software Artifacts.}
\begin{figure*}
    \centering
    \includegraphics[width=1\linewidth]{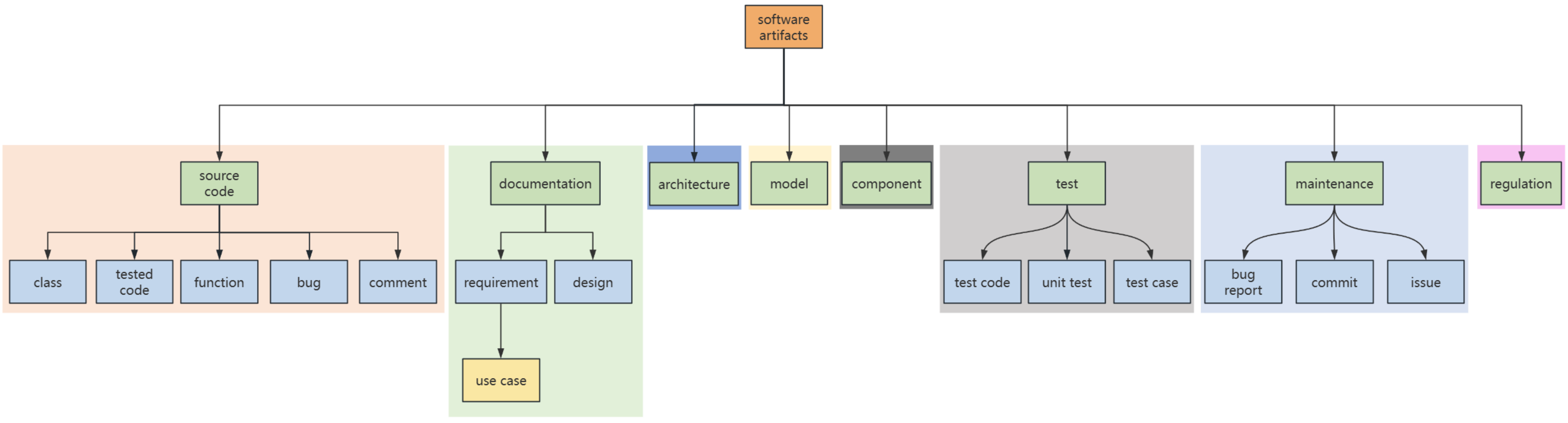}
    \caption{Hierarchical Software Artifacts Diagram. It presents a hierarchical taxonomy of 22 distinct software artifacts identified in the literature, organized into eight functional groups and three granularity layers. In each group, green, blue, and yellow nodes denote the first, second, and third hierarchical layer, respectively.}
    \label{fig:tree}
\end{figure*}

We identified 22 distinct types of software artifacts from the 76 selected papers. Based on the characteristics of these 22 artifacts and the roles they play in software systems, we classify them and construct a Hierarchical Software Artifacts Diagram, as shown in Fig. \ref{fig:tree}.
There are eight groups of analyzed software artifacts: source code artifacts, documentation artifacts, architecture artifacts, model artifacts, component artifacts, test artifacts, maintenance artifacts, and regulation artifacts.
Among them, 68\% of papers established links for ``\textit{source code}'', 41\% of papers for ``\textit{requirement}'', and 14\% of papers for ``\textit{test}''. These software artifacts represent the elements most commonly studied in the field of software traceability recovery. The majority of the research centers around ``\textit{source code}'', which is closely related to the inherent nature of software systems: the source code is frequently modified to fulfill evolving requirement and such changes often require corresponding tests updates. This dynamic interplay underscores the necessity of establishing traceability links among these artifacts.


\subsubsection{Investigated Artifacts Associations.}
\begin{figure*}
    \centering
    \includegraphics[width=0.8\linewidth]{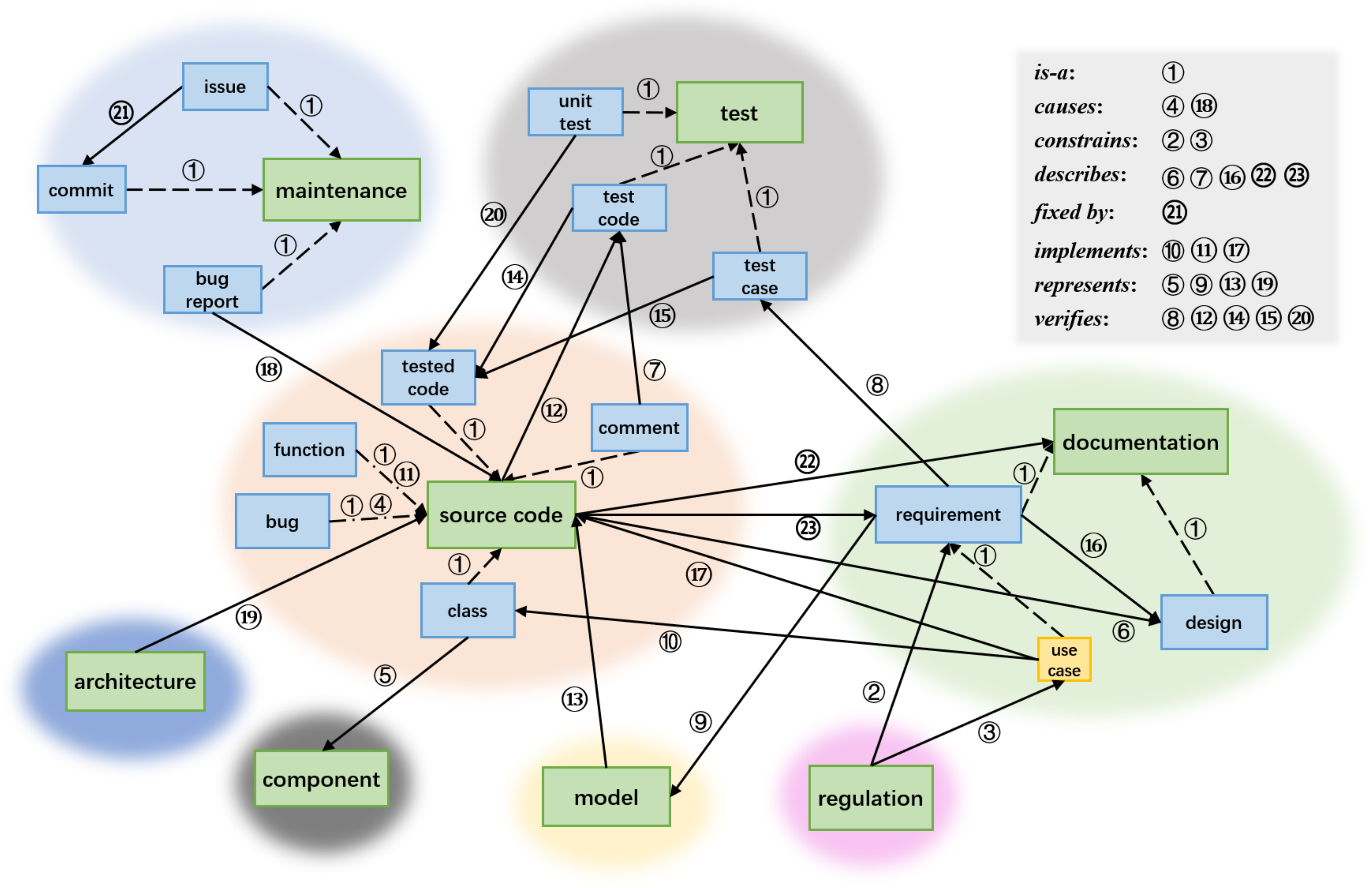}
    \caption{Artifacts Traceability Graph. Background colors represent different artifact groups in Fig. \ref{fig:tree}. The dashed lines denote ``is-a'' relationships, while solid lines represent concrete relationships. The upper-right corner displays the links corresponding to each relationship.}
    \label{fig:traceability graph}
\end{figure*}

For these studied artifacts, we found multiple types of associations resolved to build their links.
Fig. \ref{fig:traceability graph} constructs a multidimensional association network which shows the interweaving of heterogeneous relationships across different abstraction levels. These connections capture the semantic correspondence between \textbf{``describes''} and elaboration, as well as the \textbf{``implements''} mapping that reflects the evolution from abstract specifications to concrete logic. The graph also establishes paths for externally imposed \textbf{``constrains''} and for closed-loop confirmation through \textbf{``verifies''}. In addition, the association network models \textbf{``causes''} triggering relationships, \textbf{``fixed by''} mechanisms for addressing specific issues, and \textbf{``represents''} relationships across varying levels of abstraction. Together, these relationships form closed internal dependency loops within a single domain and external association structures that span across domain boundaries.

In the traceability graph, certain software artifacts appearing in multiple semantic relations indicate their multiple roles in the traceability network. Especially, \textit{``source code''} appears in multiple types of associations (e.g., \textit{``causes''}, \textit{``represents''}, \textit{``implements''}, \textit{``verifies''}, \textit{``describes''}), underscoring its central position in all phases of software development and maintenance. In addition, \textit{``requirement''} participants in \textit{``constrains''}, \textit{``represents''}, \textit{``implements''}, \textit{``verifies''}, and \textit{``describes''} traceability links, bahaving as a projection point of the semantic intent of the software. 

Traditional literature review often focus only on which artifacts are linked, which fails to uncover the significance of linking these pairs of software artifacts. To articulate the software engineering relevance of these links, we not only record the artifact pairs addressed in the papers, but also conduct an in-depth analysis of the papers, with a focus on extracting the practical problems that each link aims to solve. This helps practitioners and researchers identify high-value research directions and guides them in selecting appropriate artifact pairs for link establishment based on their needs.
In particular, for different associations among the software artifacts in Fig. \ref{fig:traceability graph}, we 
identify the significance of these associations in software engineering research. In the following, we present the internal and external associations for each group of artifacts, along with the number of involved papers.

\textbf{Associations From The Group of Source Code Artifacts.}
As the physical implementation layer of the system, the source code group embodies the concrete execution logic of the business processes. Its internal associations reflect the structured organization of code entities as follows:

\noindent \textbullet\ \circled{4} \textbf{bug -> source code (1)}: links bugs to code, which helps identify bug-prone modules and guides preventive maintenance. 

\noindent \textbullet\ \circled{11} \textbf{function -> source code (1)}: enables precise impact analysis at the function level, minimizing side effects of code updates. 

For this group of artifacts, their external associations  establish mappings between implementation artifacts and requirement, design, and testing, supporting functional verification, and change impact analysis, which includes the associations as follows:

\noindent \textbullet\ \circled{22} \textbf{source code -> documentation (9)}: lowers onboarding barriers, ensuring docs effectively help developers understand business logic. 

\noindent \textbullet\ \circled{6} \textbf{source code -> design (2)}: validates that code follows intended patterns, reducing cognitive load during maintenance. 

\noindent \textbullet\ \circled{23} \textbf{source code -> requirement (25)}: maps code to requirement to verify that the final delivery meets all original customer needs.

\noindent \textbullet\ \circled{7} \textbf{comment -> test code (1)}: explains complex test logic, reducing maintenance and reuse costs of test cases.

\noindent \textbullet\ \circled{12} \textbf{source code -> test code (1)}: maps code to tests, supporting regression optimization and ensuring stability after each commit.

\noindent \textbullet\ \circled{5} \textbf{class -> component (1)}: ensures that class implementation aligns with component logic, preventing decay and boosting reuse.

\textbf{Associations From The Group of Documentation Artifacts.}
Document artifacts define the system's business intent and technical solutions. Their internal associations reveal the evolution from requirement to design:

\noindent \textbullet\ \circled{16} \textbf{requirement -> design (3)}: ensures that every design decision is backed by a requirement, eliminating waste from over-engineering.

The external associations of this group of artifacts establish bidirectional traceability paths between \textit{requirement}, \textit{code}, and \textit{tests}, ensuring that system development aligns with stakeholder expectations:

\noindent \textbullet\ \circled{17} \textbf{use case -> source code (3)}: links user value to code, ensuring development effort is prioritized for core business features.

\noindent \textbullet\ \circled{10} \textbf{use case -> class (1)}: identifies which classes support specific user behaviors, aiding in fine-grained impact assessment.

\noindent \textbullet\ \circled{8} \textbf{requirement -> test case (1)}: ensures that all requirements are tested, directly reducing the risk of production failures due to missed logic.

\noindent \textbullet\ \circled{9} \textbf{requirement -> model (1)}: ensures that models accurately capture requirements, resolving contradictions before coding.

\textbf{Associations From The Group of Maintenance Artifacts.}
Maintenance artifacts record software evolution and repair activities. By linking change issues, defect reports, and code commits, they form a complete change log. These associations aim to improve defect localization efficiency and provide an auditable evolution history for long-term system evolution. There is one type of internal dependency and one type of external dependency:

\noindent \textbullet\ \circled{21} \textbf{issue -> commit (6)}: provides complete change logs, ensuring every edit is justified and improving collaboration.

\noindent \textbullet\ \circled{18} \textbf{bug report -> source code (4)}: speeds up bug localization, significantly shortening recovery cycles after failures.

\textbf{Associations From The Group of Test Artifacts.}
Test artifacts constitute the verification layer of quality assurance. By establishing external mappings from tests to source code artifacts, this category defines the coverage boundaries of verification activities, provides a basis for regression testing optimization, and ensures that software iterations meet quality standards.

\noindent \textbullet\ \circled{20} \textbf{unit test -> tested code (4)}: establishes feedback loops at the unit level, lowering integration debugging costs.

\noindent \textbullet\ \circled{14} \textbf{test code -> tested code (2)}: defines test coverage boundaries, enhancing safety during refactoring and preventing regressions.

\noindent \textbullet\ \circled{15} \textbf{test case -> tested code (2)}: ensures that test cases reach internal code implementations, improving bug detection efficiency.

\textbf{Associations From The Group of Regulation Artifacts.}
Regulatory artifacts represent external legal constraints and industry standards. This category transforms non-functional constraints into concrete system requirement and establishes external traceability links to support automated compliance auditing, thereby reducing legal violation risks and refactoring costs.

\noindent \textbullet\ \circled{2} \textbf{regulation -> requirement (1)}: ensures software specs adhere to industry standards, avoiding legal risks and costly refactoring.

\noindent \textbullet\ \circled{3} \textbf{regulation -> use case (1)}: ensures user scenarios do not violate industry guidelines, catching legal conflicts early.

\textbf{Associations From The Group of Model Artifacts.}
Model artifacts provide abstract representations of the system. Their external associations aim to verify whether code implementations deviate from model constraints, prevent architectural drift, and ensure that the system maintains structural robustness and consistency throughout evolution.

\noindent \textbullet\ \circled{13} \textbf{model -> source code (2)}: verifies that code has not drifted from model constraints, ensuring system robustness and scalability.

\textbf{Associations From The Group of Architecture Artifacts.}
Architecture artifacts define the system's global topology and design principles. By monitoring code compliance with architectural specifications, this category preserves system modularity and stability, preventing micro-level code changes from undermining overall system cohesion.

\noindent \textbullet\ \circled{19} \textbf{architecture -> source code (4)}: monitors if code violates architectural principles, ensuring stability during evolution or under high-load.

According to the statistical results, \textit{``source code - requirement''} emerges as the most intensively studied artifact pair, studied in 25 papers. In terms of artifact connectivity, \textit{``source code''} acts as a core node participating in the construction of 10 traceability chains, while \textit{``requirement''} are involved in 5 links; together, they form the central pillars of the traceability graph. In contrast, traceability studies involving \textit{``regulation''}, \textit{``model''}, and fine-grained entities such as \textit{``class/function''} remain relatively scarce. Approximately 45\% of the artifact pairs appear only once in the existing literature, indicating substantial research gaps in traceability for regulatory compliance, model-driven development, and low-level code entities.

\subsubsection{Survey Results}
To validate the practical value of our research, we conducted a quantitative evaluation through an online practitioner survey (results in Table \ref{tab:feedback}) and a qualitative evaluation via expert interviews (results in Table \ref{tab:expert}). 

The results of the online survey indicate strong support for our research. Among the 37 valid responses, we calculated the proportion of participants who assigned scores of 4 or 5 to each question. 92\% of participants recognized the clarity, rationality, and practical value of our artifacts traceability graph, and 95\% acknowledged its completeness. In addition, 92\% of participants affirmed the interpretability of our software artifact hierarchy, and 95\% recognized its completeness.

\begin{table}[]
\caption{Feedback Results of Online Survey.}
\label{tab:feedback}
\centering
\resizebox{0.5\textwidth}{!}{
\begin{tabular}{cccc}
\hline
\# Feedback & \# Valid Feedback & Artifacts Completeness & Hierarchy Clarity \\ \hline
38 & 37 & 95\% & 92\% \\ \hline \hline
Association Completeness & Associations Clarity & Associations Value     & Associations Rationality \\ \hline
95\% & 92\% & 92\% & 92\% \\ \hline
\end{tabular}
}
\end{table}

A qualitative evaluation conducted through expert interviews (n = 4) further validated the values of our software artifact relationship network. 
Two industry experts emphasized the study's potential for addressing real engineering challenges, noting that it provides a clear practical pathway for managing architectural evolution risks and optimizing automated testing workflows.
Meanwhile, two academic experts valued the study's contribution to decomposing complex semantic relationships and identifying research gaps in the field, considering it a solid theoretical foundation for future work on LLM-driven traceability and automated compliance auditing.
This dual validation confirms that our study not only systematizes knowledge at the theoretical level, but also demonstrates practical value for guiding real-world tasks.

\begin{table*}[]
\caption{Responses of Expert Interview.}
\label{tab:expert}
\centering
\resizebox{0.95\textwidth}{!}{
\begin{tabular}{lll}
\hline
\multicolumn{3}{l}{\begin{tabular}[c]{@{}l@{}}\textbf{Expert 1}: I work on architecture maintenance, and my biggest concern is architectural drift - a small code change can easily \\ push the architecture off track. This study points out that research on the architecture -> source code link is extremely limited, \\ which really hits the pain point. It made me realize we shouldn't focus only on requirements; architecture   should be treated \\ as the top-level artifact, with automated checks to keep the system cohesive.\end{tabular}} \\ \hline
\multicolumn{3}{l}{\begin{tabular}[c]{@{}l@{}}\textbf{Expert 2}: I'm responsible for CI/CD pipelines and constantly deal with bloated test suites. The paper highlights that source \\ code is the hub of ten traceability chains, and it distinguishes between test cases and test code, which is very practical. \\ This inspired me to anchor traceability on code and use   bidirectional links to prune useless test cases from the pipeline.\end{tabular}} \\ \hline
\multicolumn{3}{l}{\begin{tabular}[c]{@{}l@{}}\textbf{Expert 3}: I use LLMs for automated traceability. I used to think traceability was just a simple binary matching problem. But \\ your idea of a multidimensional association network was quite enlightening. Treating requirements as a semantic \\ projection point and distinguishing different relation weights made me realize that we shouldn't treat all links the \\ same - I need to design more fine-grained prompts for LLMs.\end{tabular}} \\ \hline
\multicolumn{3}{l}{\begin{tabular}[c]{@{}l@{}}\textbf{Expert 4}: I research compliance auditing for high-assurance systems, which has long been a niche area. The statistics in \\ this study are quite revealing: regulation-related links are extremely scarce, and nearly half of the artifact pairs appear \\ only once. The paths you outlined, such as Regulation -> Requirement, basically provide the backbone for my research \\ and strengthen my belief in using traceability graphs as evidence chains for compliance.\end{tabular}} \\ \hline
\end{tabular}
}
\end{table*}

\subsubsection{Guidelines}
Based on our findings, we propose three guidelines to advance artifacts traceability research, which provides a systematic roadmap for software management.

\textbf{Multi-hop Traceability Chains.}
While existing studies predominantly focus on specific binary artifact pairs (e.g., ``\textit{requirement - source code}''), these isolated relationships are often insufficient for complex engineering decisions. We observe that many critical but missing associations can be bridged by integrating multi-hop traceability chains \cite{72}. For example, by leveraging an intermediary artifact $C$ to connect $A$ and $B$ (where $A \to C$ and $C \to B$ links can be well established), we can construct a global association network. This shift from localized binary mappings to path-based traceability provides a systematic guideline for uncovering hidden dependencies among heterogeneous artifacts.

This chain-based recovery strategy reveals deep relationships that remain invisible to single-pair methods, supporting more sophisticated analytical tasks. Following this guideline, Table \ref{table:RQ1-chain} presents four highly valuable artifact pairs for which no direct recovery methods were proposed in our literature pool. For each pair, we propose indirect traceability paths by utilizing intermediate artifacts as stepping stones, through which their associations can be effectively established and explored in future research. For example, a traceability chain \textit{``bug report''}->\textit{``source code''}->\textit{``requirement''} illustrates the interconnected process of bug finding and user requirement. Such chains facilitate establishing potential links between artifacts within the sequence via intermediate associations. If there is a significant gap between two artifacts in the chain, appropriate intermediate artifacts can be chosen to provide additional semantic information, thereby extending the scope of link establishment.

\begin{table*}[]
\caption{Examples of Proposed Indirect Traceability Chains for Latent Artifact Associations.}
\resizebox{\textwidth}{!}{
\begin{tabular}{|l|l|l|}
\hline
\multicolumn{1}{|c|}{\textbf{Missing Artifacts Associations}} & \multicolumn{1}{c|}{\textbf{Research Value}} & \multicolumn{1}{c|}{\textbf{Multi-hop Traceability Chain}} \\ \hline

regulation -> test case & \begin{tabular}[c]{@{}l@{}}It maps legal constraints to verifiable entities, which enables automated \\ compliance auditing and ensures that system behavior conforms to regulatory \\ standards with empirical evidence.\end{tabular} & \textbf{regulation} -> requirement -> \textbf{test case} \\ \hline

bug report -> requirement & \begin{tabular}[c]{@{}l@{}}It links defects to requirement distinguishes coding errors from requirement \\ ambiguities. By identifying requirement characteristics that induce failures, \\ this approach improves requirement engineering quality and reduces rework costs.\end{tabular} & \textbf{bug report} -> source code -> \textbf{requirement} \\ \hline

architecture -> requirement & \begin{tabular}[c]{@{}l@{}}It 
 analyzes the alignment between implementation and business objectives to \\ identify architectural erosion. Quantifying architectural support for functional \\ evolution provides a quantitative basis for refactoring decisions.\end{tabular} & \textbf{architecture} -> source code -> \textbf{requirement} \\ \hline

model -> test code & \begin{tabular}[c]{@{}l@{}}It establishes alignment between design specifications and test logic, ensuring that \\ verification conforms to architectural intent. It supports automatic derivation of \\ test scripts when models change, improving the level of automated verification.\end{tabular} & \textbf{model} -> source code -> \textbf{test code} \\ \hline
\end{tabular}
}
\label{table:RQ1-chain}
\end{table*}

\textbf{The Central Pivot (requirement - Code).} In the landscape of software evolution, the consistency between \textit{source code} and \textit{requirement} forms the backbone of traceability research. This link is more than a binary association; it is the essential semantic bridge between business intent (``what'') and technical implementation (``how''). By anchoring high-level requirement to low-level source code, this relationship provides the most direct evidence for verifying functional completeness and ensuring that the final delivery aligns with stakeholder expectations.

From a structural perspective, the ``\textit{source code - requirement}'' link exhibits extraordinary topological centrality within the global Artifacts Traceability Graph. Our analysis reveals that these two artifacts serve as the primary hubs for nearly all long-range traceability paths: source code mediates links to architecture, tests, and defects, while requirement anchor upstream regulation and downstream design. If this core consistency is compromised, the global graph fragments into isolated ``knowledge islands''.

Consequently, we propose a synchronous construction guideline: traceability should shift from post-hoc recovery to early-stage integration. Establishing bidirectional ``\textit{code - requirement}'' links from the project's inception creates a ``single source of truth''. This mechanism not only enables precise change impact analysis and optimized regression testing, but also secures an immutable evidence chain for compliance.

\textbf{The Quality Boundaries (Peripheral Links).} In contrast to core artifacts, peripheral artifact pairs (such as ``\textit{regulation - use case}'', ``\textit{class - component}'', and ``\textit{requirement - test case}'') are located at the margins of the traceability graph. These links connect different stages of the software lifecycle and involve cooperation between different experts (e.g., architects and testers). Neglecting these marginal relationships could lead to practical problems in software engineering, for example:

\noindent \textbullet\ \textbf{neglecting ``\textit{regulation - use case}''}: The system may function correctly but fail to meet security standards, leading to high rework costs or legal risks when violations are discovered late.

\noindent \textbullet\ \textbf{neglecting ``\textit{class - component}''}: Frequent code changes without architectural alignment make the system messy, hindering future updates or technology migrations.

\noindent \textbullet\ \textbf{neglecting ``\textit{requirement - test case}''}: It is unclear whether core logic is actually verified, leading to redundant testing of minor features while critical risks remain hidden.

\noindent \textbullet\ \textbf{neglecting ``\textit{model - requirement}''}: When technical models drift from business intent, the resulting system may be technically functional but fail to solve the actual business problem.

Researching these peripheral links is essential for a full-lifecycle management approach. Moving beyond simple point-to-point traceability to explore these less-studied links helps build a self-explanatory software system. In such an environment, every piece of code and every test can be traced back to its business purpose and legal origin, ensuring the software is reliable and easy to maintain over the long term.

\begin{mdframed}[style=graystyle]
\textbf{Answer to RQ1:} {There are 22 types of software
artifacts and 23 types of associations analyzed in the current research. Half of existing studies recovered the links between different documentations and the code they describe, which is the biggest research hotspot.}
\end{mdframed}

\begin{table*}[]
\caption{Summary of Techniques, Representations, and Evaluation Methods for Artifact Linking Tools}
\resizebox{\textwidth}{!}{
\begin{tabular}{|l|l|l|r|r|l|}
\hline
\multicolumn{1}{|c|}{\textbf{Artifacts Pair (\#Papers)}} & \multicolumn{1}{c|}{\textbf{Technique}} & \multicolumn{1}{c|}{\textbf{Input Representation}} & \multicolumn{1}{c|}{\textbf{\begin{tabular}[c]{@{}c@{}}Code \\ Public\end{tabular}}} & \multicolumn{1}{c|}{\textbf{\begin{tabular}[c]{@{}c@{}}Dataset \\ Public\end{tabular}}} & \multicolumn{1}{c|}{\textbf{Metrics}} \\ \hline

source code - requirement (25) & \begin{tabular}[c]{@{}l@{}}IR, DFA, PA, \\ Eye-Tracking, \\ Manual, LLM\end{tabular} & \begin{tabular}[c]{@{}l@{}}\textbf{source code:} AST, bytecode, token, \\ events, XML, identifier, set, \\ dynamic call graph, structured code, \\ structured text, file, call graph\\ \textbf{requirement:} text, ID-RTM, RTM, \\ labels, issues, structured text, identifier\end{tabular} & 11/25 & 25/25 & \begin{tabular}[c]{@{}l@{}}Recall, Precision, AP, MAP, \\ IGR, F-Measure, FP, Cliff's delta, \\ P-value, DiffAR, Effort, \\ Selection Proportion, Correctness, \\ Incorrectness, Similarity, \\ Conflicts, Pass@1, Coverage\end{tabular} \\ \hline

source code - documentation (9) & \begin{tabular}[c]{@{}l@{}}IR, ML, PA, \\ Matching\end{tabular} & \begin{tabular}[c]{@{}l@{}}\textbf{source code:} term, identifier, text, file, \\ call graph, set o words\\ \textbf{documentation:} text, file, model\end{tabular} & 1/9 & 8/9 & Recall, Precision, F-Measure, REI \\ \hline

issue - commit (6) & \begin{tabular}[c]{@{}l@{}}IR, ML, DL, \\ CL, PLM\end{tabular} & \begin{tabular}[c]{@{}l@{}}\textbf{issue:} text, structured text, metadata\\ \textbf{commit:} text, code, structured text, \\ metadata\end{tabular} & 4/6 & 6/6 & \begin{tabular}[c]{@{}l@{}}Recall, Precision, AP, MAP, AUC, \\ MCC, PF, MRR, ACC, F-Measure, \\ P-value, Hit\end{tabular} \\ \hline

unit test - tested code (4) & \begin{tabular}[c]{@{}l@{}}DL, Matching, \\ DFA, PA, PLM\end{tabular} & \begin{tabular}[c]{@{}l@{}}\textbf{unit test:} text, token, vector, code\\ \textbf{tested code:} text, identifier, vector, code\end{tabular} & 1/4 & 4/4 & \begin{tabular}[c]{@{}l@{}}Recall, Precision, ACC, F-Measure, \\ Cliff's delta, P-value\end{tabular} \\ \hline

architecture - source code (4) & \begin{tabular}[c]{@{}l@{}}IR, ML, DFA, \\ Matching, PA\end{tabular} & \begin{tabular}[c]{@{}l@{}}\textbf{architecture:} keywords, text, identifier, \\ label\\ \textbf{source code:} term, text, identifier, vector\end{tabular} & 2/4 & 4/4 & \begin{tabular}[c]{@{}l@{}}Recall, Precision, ACC, \\ F-Measure, Correctness, \\ Effectiveness, Specificity\end{tabular} \\ \hline

bug report - source code (4) & \begin{tabular}[c]{@{}l@{}}IR, ML, \\ Matching\end{tabular} & \begin{tabular}[c]{@{}l@{}}\textbf{bug report:} text\\ \textbf{source code:} file, entity, identifier, AST\end{tabular} & 3/4 & 4/4 & \begin{tabular}[c]{@{}l@{}}Recall, Precision, MAP, MRR, Hit, \\ Usefulness, Effectiveness\end{tabular} \\ \hline

use case - source code (3) & \begin{tabular}[c]{@{}l@{}}IR, ML, PA, \\ Matching\end{tabular} & \begin{tabular}[c]{@{}l@{}}\textbf{use case:} text, label\\ \textbf{source code:} text, event, file\end{tabular} & 0/3 & 2/3 & \begin{tabular}[c]{@{}l@{}}Recall, Precision, VPR, ACC, \\ Cliff's delta, P-value\end{tabular} \\ \hline

requirement - design (3) & IR, DL & \begin{tabular}[c]{@{}l@{}}\textbf{requirement:} text, structured text\\ \textbf{design:} model element, text\end{tabular} & 1/3 & 1/3 & Recall, Precision, MAP, F-Measure \\ \hline

test case - tested code (2) & \begin{tabular}[c]{@{}l@{}}Matching, \\ Manual\end{tabular} & \begin{tabular}[c]{@{}l@{}}\textbf{test case:} text, XML\\ \textbf{tested code:} code, XML\end{tabular} & 1/2 & 2/2 & Counts, Graph Connectivity \\ \hline

test code - tested code (2) & PA & \begin{tabular}[c]{@{}l@{}}\textbf{test code:} structured code, \\ runtime execution\\ \textbf{tested code:} structured code, bytecode\end{tabular} & 1/2 & 2/2 & \begin{tabular}[c]{@{}l@{}}Recall, Precision, MAP, AUC, \\ F-Measure\end{tabular} \\ \hline

model - source code (2) & IR, DFA & \begin{tabular}[c]{@{}l@{}}\textbf{model:} identifier, structured text\\ \textbf{source code:} vector, structured code\end{tabular} & 0/2 & 0/2 & \begin{tabular}[c]{@{}l@{}}Recall, Precision, FP, Effort, \\ Effectiveness\end{tabular} \\ \hline

source code - test code (1) & IR & \begin{tabular}[c]{@{}l@{}}\textbf{source code:} token, AST\\ \textbf{test code:} token, AST\end{tabular} & 1/1 & 1/1 & \begin{tabular}[c]{@{}l@{}}Recall, Precision, AUC, ACC, \\ F-Measure\end{tabular} \\ \hline

function - source code (1) & PA & \begin{tabular}[c]{@{}l@{}}\textbf{function:} list\\ \textbf{source code:} file\end{tabular} & 0/1 & 1/1 & Feasibility, Coverage \\ \hline

source code - UML (1) & Eye-Tracking & \begin{tabular}[c]{@{}l@{}}\textbf{source code:} visual widgets\\ \textbf{UML:} visual widgets\end{tabular} & 0/1 & 0/1 & - \\ \hline

use case - class (1) & IR & \begin{tabular}[c]{@{}l@{}}\textbf{use case:} text\\ \textbf{class:} structured code\end{tabular} & 0/1 & 1/1 & \begin{tabular}[c]{@{}l@{}}Recall, Precision, AP, MAP, \\ F-Measure, Cliff's delta, P-value\end{tabular} \\ \hline

requirement - model (1) & IR & \begin{tabular}[c]{@{}l@{}}\textbf{requirement:} text\\ \textbf{model:} element, text\end{tabular} & 1/1 & 1/1 & \begin{tabular}[c]{@{}l@{}}Recall, Precision, AUC, MCC, \\ F-Measure\end{tabular} \\ \hline

requirement - test case (1) & IR, ML & \begin{tabular}[c]{@{}l@{}}\textbf{requirement:} feature vector\\ \textbf{test case:} feature vector\end{tabular} & 0/1 & 1/1 & Recall, Precision \\ \hline

comment - test code (1) & IR, PA & \begin{tabular}[c]{@{}l@{}}\textbf{comment:} text\\ \textbf{test code:} code snippets\end{tabular} & 1/1 & 1/1 & Recall, Precision, F-Measure \\ \hline

source code - design (1) & Matching & \begin{tabular}[c]{@{}l@{}}\textbf{source code:} XML\\ \textbf{design:} UML\end{tabular} & 0/1 & 1/1 & Intersection, Agreements \\ \hline

class - component (1) & Manual & \begin{tabular}[c]{@{}l@{}}\textbf{class:} UML\\ \textbf{component:} UML\end{tabular} & 0/1 & 1/1 & Effectiveness \\ \hline

bug - source code (1) & DFA & \begin{tabular}[c]{@{}l@{}}\textbf{bug:} patch\\ \textbf{source code:} file\end{tabular} & 0/1 & 1/1 & Agreements, Disagreements \\ \hline

use case - regulation (1) & IR & \begin{tabular}[c]{@{}l@{}}\textbf{use case:} text\\ \textbf{regulation:} text\end{tabular} & 0/1 & 0/1 & AP, MAP, ACC \\ \hline

regulation - requirement (1) & IR, ML & \begin{tabular}[c]{@{}l@{}}\textbf{regulation:} text\\ \textbf{requirement:} text\end{tabular} & 0/1 & 1/1 & Recall, Precision, AP, F-Measure \\ \hline
\end{tabular}
}
\begin{flushleft}
{\footnotesize Values in the ``Code Public'' and ``Dataset Public'' columns denote the number of papers releasing public resources out of the total papers for that task.}
\end{flushleft}
\label{table:RQ2}
\end{table*}

\subsection{(RQ2) Current Status of Linking Tools}
We investigated the existing linking tools by analyzing the representations of artifacts, the techniques used, and the evaluation designs, which is reported in Table \ref{table:RQ2}. 

\subsubsection{Artifact Input Reprsentation.}
Certain linking techniques can only be applied on specific representations of artifacts. We compiled a list of artifact representations identified in the reviewed literature shown in the third column of Table \ref{table:RQ2}. Basically, these various artifact representations can be classified into five categories: 

\noindent \textbullet\ \textbf{Linguistic Textual Representations:} This category primarily consists of human-written text or discrete linguistic units that lack programming syntax constraints. 16 software artifacts use this representation; e.g., ``\textit{requirement}'', ``\textit{bug report}'', and `\textit{issue}'' can be represented by text or structured text. 

\noindent \textbullet\ \textbf{Static Implementation Structures:} It represents implementation logic and structural features governed by formal programming language syntax or static rules. 12 artifacts use this representation; e.g., ``\textit{source code}'' and ``\textit{test code}'' can be represented by AST, code snippets, or structured code.

\noindent \textbullet\ \textbf{High-Level Design Blueprints:} This category focuses on high-level system architecture, logical relationships, or interface layouts while abstracting away specific implementation details. 5 software artifacts use this representation; e.g., \textit{class} is represented by UML, \textit{design} by UML or model element.

\noindent \textbullet\ \textbf{Dynamic Runtime Behaviors:} It captures runtime behaviors and temporal sequences of a program during its actual or simulated execution. 3 software artifacts use this representation; e.g., \textit{test code} is represented by runtime execution.

\noindent \textbullet\ \textbf{Mathematical Relational Models Representations:} This category transforms software information into mathematical objects or relational matrices to facilitate computational analysis and link recording. 6 software artifacts use this representation;, e.g., ``\textit{commit}'' is represented by metadata.


\subsubsection{Linking Technique}
Through an in-depth analysis of proposed tools, the techniques used for each pair of artifacts are presented in the second column of Table \ref{table:RQ2}. The linking techniques employed between artifacts are typically limited to only one or two categories. But the link of ``\textit{requirement}'' and ``\textit{source code}'' can apply nearly multiple techniques. 
We summarize the following categories of linking techniques:

\noindent \textbullet\ \textbf{Information Retrieval (IR):} IR is the most dominant technique, accounting for as much as 57\% of papers. It relies on textual similarity methods including TF-IDF and VSM to connect artifacts such as requirement and models \cite{7}.

\noindent \textbullet\ \textbf{Static/Dynamic Program Analysis (PA):} Used in 21 studies, these program analysis techniques establish links between artifacts based on structural dependencies or runtime behaviors, such as call graphs \cite{49} or code coverage \cite{9}. 

\noindent \textbullet\ \textbf{Machine Learning (ML):} ML is an emerging trend (10 studies). It trains models to recognize complex patterns to link artifacts by integrating various data sources \cite{6,22,25,29,32,40,41,43,45}. 

\noindent \textbullet\ \textbf{Rules/Pattern Matching (Matching): }Widely used in 7 studies, it uses predefined rules or consistent patterns to create links between artifacts; e.g., linking a function with the test whose name includes the function name \cite{18,34,35}. 

\noindent \textbullet\ \textbf{Data Flow Analysis (DFA): }Used in 6 studies, DFA tracks the transformation of data across variables and logical paths within a program. It is primarily employed to identify functional dependencies between artifacts or to verify the logical consistency of traceability links by analyzing how information propagates through the code implementation \cite{3,14,30,36,63,64}.

\noindent \textbullet\ \textbf{Eye-Tracking}: Used in 3 studies, this technique involves monitoring developers' visual attention patterns while they navigate between artifacts. As a non-automated approach specifically focused on connections between \textit{source code} and other software artifacts, it is primarily used to explore cognitive processes or manually validate the quality of established links \cite{12,26,52}.

\noindent \textbullet\ \textbf{Manual Establishment (Manual):} Reported in 3 studies, manual traceability involves human experts establishing or verifying links. As a non-automated approach that is generally not categorized as an experimental method, it is also dedicated to evaluating the performance of other automated techniques \cite{4,17,58}.

\noindent \textbullet\ \textbf{Deep Learning (DL):} Used in 3 studies, DL leverages neural architectures such as CNNs and RNNs to automatically extract hierarchical features from artifacts, capturing deep semantic relationships that traditional IR might miss \cite{21,34,38}.

\noindent \textbullet\ \textbf{Pre-trained Language Models (PLM):} Featured in 3 latest studies, PLMs such as CodeBERT or GraphCodeBERT utilize large-scale pre-training on code and natural language corpora to provide context-aware embeddings, serving as a powerful backbone for modern traceability tasks \cite{34,38,39}.

\noindent \textbullet\ \textbf{Large Language Model (LLM):} As a new and promising direction (1 studies), LLMs such as GPT \cite{achiam2023gpt} rely on advanced semantic understanding to connect artifacts across different languages and abstraction levels, such as between \textit{requirement} and \textit{source code} \cite{55}.

\noindent \textbullet\ \textbf{Contrastive Learning (CL):} Employed in 1 studies, CL is an advanced subset of representation learning that trains models to pull related artifact pairs closer in the embedding space while pushing unrelated ones apart, significantly improving the accuracy of cross-modal retrieval \cite{39}.

Among these categories of techniques, traditional methods, such as \textit{IR}, \textit{Matching}, \textit{DFA}, and \textit{PA}, offer several advantages. They are well-established with a wealth of mature research available that enables practitioners to select suitable methods based on specific needs. They typically require fewer computational resources and hardware support.
However, these traditional approaches have weak semantic understanding and are only suitable for well-structured and standard-design softwares. In addition, \textit{Manual Establishment} and \textit{Eye-Tracking} rely on human effort to establish links, which are generally not used in routine experiments.

To overcome these shortcomings, more recent and advanced techniques, including \textit{LLM}, \textit{PLM}, \textit{DL}, \textit{ML}, and \textit{CL}, demonstrate significantly stronger semantic comprehension for large-scale and complex softwares.
However, they require substantial resources for training models and constructing datasets. 
Furthermore, their application in this domain is still in its early stages, and achieving optimal performance often necessitates extensive customization, experimentation, and research.



\subsubsection{Evaluation}
We analyzed three aspects of evaluation designs used in the literature, reported in the last three columns of Table \ref{table:RQ2}.

\textbf{Code Accessibility.}
Only 37\% of selected studies have released their technical source code. These open source implementations are mainly concentrated in studies that focus on traceability between \textit{source code} and \textit{documentation}. This trend may be attributed to the adoption of well-established IR and NLP techniques. They provide a valuable baseline for future research, allowing for reproducibility, comparison, and further extension.
In contrast, the linking techniques used in the remaining studies are either non-reproducible or difficult to reproduce. This means that a majority of contributions in this field cannot be easily validated by the research community, which is a big challenge for fair and consistent comparisons.

\textbf{Dataset Accessibility.} Compared with code accessibility, dataset availability is relatively higher: 89\% of the selected papers made their datasets available. Some papers used a combination of public benchmark datasets, modern open-source projects, and industrial/private datasets; we regard such cases as having disclosed the datasets used in the paper.


\textbf{Metrics.}
A variety of evaluation metrics were found in evaluating the performance of different linking tools. 
The task of software traceability link recovery is fundamentally framed as either a binary classification problem - determining whether a potential link is a ``true'' or ``false'' link - or as an IR problem - retrieving a set of relevant target artifacts for a given source artifact. To evaluate solutions to such problems, we found three commonly used metrics: \textit{Recall}, \textit{Precision}, and \textit{F}, which used in 67\%, 68\%, and 43\% of research articles, respectively. These commonly adopted metrics can provide guidance for the evaluation experiments conducted in future research, yet other metrics can also be chosen based on the specific artifacts involved (e.g., using \textit{Coverage} to assess established links between \textit{test cases} and \textit{source code}).

\subsubsection{Guidelines}
To move beyond a simple list of tools, we synthesize the identified techniques into a practical framework for researchers. This subsection provides representation-technique-cost decision support to help researchers choose tools that fit their available resources. Furthermore, we propose common datasets and minimum metrics criteria to reduce experimental bias and ensure the reliability of future evaluations. Through this, we offer a holistic framework that bridges the gap between theoretical techniques and experimental rigor.

\begin{figure*}
    \centering
    \includegraphics[width=1\linewidth]{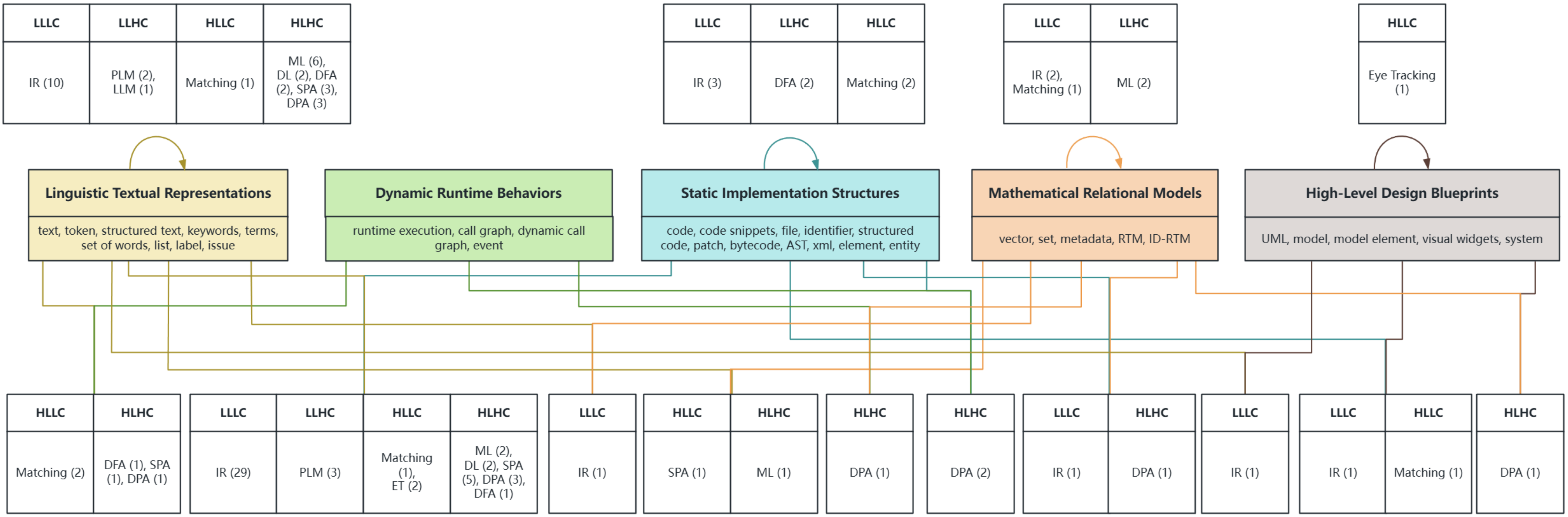}
    \caption{Technical Decision Map for Artifacts Traceability Linking. The techniques for each pair of artifact representations are categorized into four groups based on their costs: LLLC (low labor and low computation), LLHC (low labor and high computation), HLLC (high labor and low computation), and HLHC (high labor and high computation).}
    \label{fig:representation-technique-cost}
\end{figure*}

\textbf{Technical Decision Map for Traceability Recovery.}
We argue that establishing traceability links essentially entails computing semantic or structural similarity between artifact representations via various automated techniques. However, the selection of specific recovery approaches is not arbitrary; instead, it is contingent upon intrinsic properties of the artifacts and available resource constraints. We further acknowledge that different techniques entail varying levels of human effort and computational overhead. To offer actionable guidance to researchers on selecting suitable recovery approaches, we propose a technical decision map shown in Fig. \ref{fig:representation-technique-cost}, where we further analyzed all papers and mapped five categories of artifact representations to the corresponding recovery techniques and their associated cost profiles.

Our analysis categorizes these techniques based on their cost profiles, ranging from LLLC (Low Labor, Low Computation) to HLHC (High Labor, High Computation). The detailed definition criteria are presented in Table \ref{tab:cost}. Labor cost was determined based on whether the research process entailed manual tasks such as data labeling, dataset construction, or groundtruth creation; while computation cost was determined based on whether a study's experimental settings involve high-performance servers, specialized hardware resources, or high computational intensity.

\begin{table*}[]
\caption{Definition Criteria for the Cost of Linking Techniques}
\label{tab:cost}
\resizebox{\textwidth}{!}{
\begin{tabular}{|c|l|l|}
\hline
\textbf{Dimension} & \multicolumn{1}{c|}{\textbf{Low Definition Criteria}} & \multicolumn{1}{c|}{\textbf{High Definition Criteria}} \\ \hline
\textbf{Labor} & \begin{tabular}[c]{@{}l@{}}\textbf{Low Labor (LL):} No manual standard data is required; datasets and groundtruth are not \\ manually created, directly using existing public benchmarks.\end{tabular} & \begin{tabular}[c]{@{}l@{}}\textbf{High Labor (HL):} Labor-intensive tasks are required (e.g., manual labeling, manual dataset/groundtruth construction, \\ and manual validation).\end{tabular} \\ \hline
\textbf{Computation} & \begin{tabular}[c]{@{}l@{}}\textbf{Low Computation (LC):} Only basic computational resources are required (e.g., standard PCs); \\ the algorithm has low complexity and does not require specialized hardware acceleration.\end{tabular} & \begin{tabular}[c]{@{}l@{}}\textbf{High Computation (HC):} The implementation requires sustained computational power from high-performance servers, \\ specialized hardware such as GPUs, or experimental setups with high computational intensity.\end{tabular} \\ \hline
\end{tabular}
}
\end{table*}

The technical decision map in Fig. \ref{fig:representation-technique-cost} illustrates a sophisticated trade-off between the advancement of linking techniques and the associated recovery costs. Our analysis reveals that links between natural language and source code exhibit the highest diversity in techniques, ranging from low-cost IR methods to computationally intensive ML/DL models. Notably, HLHC techniques are predominantly concentrated in scenarios requiring the bridging of significant semantic gaps, such as mapping mathematical abstractions or design models to dynamic execution traces. This pattern suggests that while automation is advancing, researchers must still navigate the tension between precision and resource investment when dealing with high-level or volatile artifacts.

\textbf{Common Datasets For Evaluation.}
Datasets are a key factor in experimental evaluation, as their quality and characteristics directly influence the performance and generalizability of traceability recovery techniques. Our analysis categorizes the datasets used in the literature into three types:
\textit{(a) Benchmark Datasets (used in 25\% of papers):} widely cited and highly comparable, yet often suffer from being outdated, small-scale, and poorly maintained;
\textit{(b) Selected Open-Source Projects (used in 74\% of papers):} large-scale projects sourced from platforms such as GitHub and Jira. They reflect contemporary development practices but may lack unified ground truth;
\textit{(c) Industrial/Private Datasets (used in 24\% of papers):} offer high realism but generally inaccessible and irreproducible for the broader research community.

\begin{figure*}
    \centering
    \includegraphics[width=0.9\linewidth]{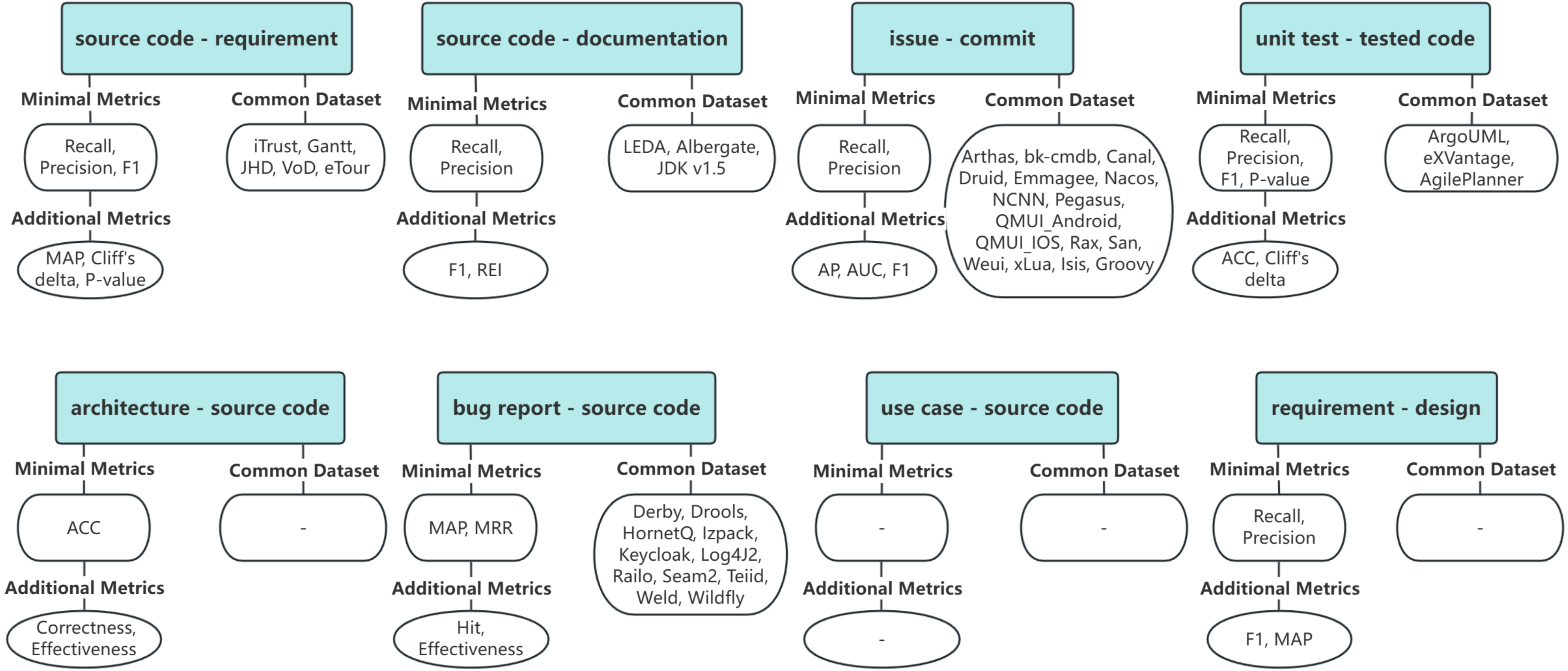}
    \caption{Metrics and Datasets Recommendation for Software Artifact Pairs. This figure summarizes common datasets and minimum/additional metrics criteria for artifact pairs studied in the literature.}
    \label{fig:metrics and datasets}
\end{figure*}

To mitigate performance biases stemming from heterogeneous dataset selection, we propose sets of ``common datasets'' for artifact pairs studied in more than two papers. We define common datasets as those used in over 50\% of papers or over five papers in a pair category. 
As illustrated in Fig. \ref{fig:metrics and datasets}, specific datasets such as iTrust for ``\textit{source code} - \textit{requirement}'' links and JDK v1.5 for ``\textit{source code} - \textit{documentation}'' links, serve as established benchmarks. We strongly recommend that researchers prioritize these common datasets to ensure results are comparable across different studies.

Notably, our analysis also reveals a lack of common datasets for emerging or complex artifact pairs, such as ``\textit{architecture} - \textit{source code} and \textit{use case} - \textit{source code}''. This absence of standardized benchmarks in these areas hinders the direct comparison of different recovery techniques and suggests a critical need for the community to develop and share high-quality, open-source ground truth data for these specific scenarios.

\textbf{Minimum Metrics Criteria For Evaluation.}
We observed that the evaluation metrics are not entirely consistent across different pairs of artifacts. To improve the reliability and comparability of the experimental results, we also extracted a minimum metrics criterion for artifact pairs studied in more than two papers.

Similar to our dataset selection logic, a metric is included in the minimum metrics criteria if its usage exceeds 50\% in its category, while those between 20\% and 50\% are labeled as additional metrics. 
As illustrated in Fig. \ref{fig:metrics and datasets}, the selection of metrics is highly sensitive to the nature of the traceability task. For requirement-related links, \textit{Recall}, \textit{Precision}, and \textit{F1-score} constitute the minimum reporting requirement, ensuring a balanced evaluation of retrieval completeness and accuracy. In contrast, for tasks involving bug reports or architecture abstractions, ranking-oriented metrics such as MAP and MRR are prioritized to reflect the effectiveness of automated recommendation lists. The absence of recommended metrics for ``\textit{use case} - \textit{source code}'' still highlights a lack of community consensus.
Adhering to these criteria ensures that the most critical information of interest for a given artifact pair is consistently evaluated.


\begin{mdframed}[style=graystyle]
\textbf{Answer to RQ2:} {The current landscape of traceability tools is dominated by IR (57\%), though a shift toward resource-intensive semantic models (especially ML/DL and PLM/LLM) is emerging to bridge complex artifact gaps. While dataset accessibility is high (89\%), the community faces significant challenges in reproducibility due to low code transparency (37\%) and a lack of standardized benchmarks for high-level abstractions.}
\end{mdframed}

\subsection{(RQ3) Usage Scenarios}
To understand the application of recovered links, we investigated their domains, involved software phases, and objectives.

\subsubsection{Domain}
The usage scenarios of different artifact associations are summarized in Table \ref{table:RQ3}. We found that 95\% of linking approaches were applied in an academic setting and 5\% were applied in specific industrial projects. Among them, five papers addressed both academic and industrial concerns. 72 academic-oriented studies covered a wide range of software artifacts, whereas 4 industry-oriented studies focused on a few pairs of artifacts that receive little attention in academic research. 
For example, Bonner et al. \cite{19} recovered the links between \textit{design} and \textit{requirement} in the domain of automotive electronics and electrical systems development. 

\subsubsection{Software Lifecycle Phase}
The fourth column of Table \ref{table:RQ3} shows that the traceability links between artifacts studied by existing research participate in nearly all phases of the software lifecycle. Half of existing studies aimed at restoring links appeared in the ``implementation'' phase, followed by the ``maintenance'' phase which was targeted by 26\% of studies. In long-lived software systems that undergo multiple development and maintenance cycles, the ``implementation'' phase determines the creation and quality of traceability links, while the ``maintenance'' phase determines their longevity and value.

\subsubsection{Objective}
We categorized the objectives of the recovered links between artifacts in all collected papers.

\noindent \textbullet\ \textbf{Requirements-Implementation Consistency:}  In this category, artifact links are mainly recovered for maintaining consistency between \textit{requirement} and \textit{code}. For example, Mona et al. \cite{60} addressed the problem that traceability links rapidly degrade due to code refactoring during software evolution, leading to a loss of consistency between requirement documents and implementation code.



\noindent \textbullet\ \textbf{Quality Assurance and Maintenance:} The primary focus within this category is to support defect localization. For example, Zhang et al. \cite{39} proposed an efficient pretraining framework named ``EALink'', which addresses the problem that missing associations between \textit{issue} and \textit{commit} in software maintenance adversely affect defect localization and prediction accuracy.


\noindent \textbullet\ \textbf{Program Comprehension:} This category aims to facilitate program comprehension by building links between documentation and implementation.
For example, Nepomuceno et al. \cite{4} validated the effectiveness of the SMarty traceability mechanism to assist developers in understanding the impact of configurations across versions and diagrams.


\noindent \textbullet\ \textbf{Model-driven System Understanding:} The most extensively studied objective is to maintain consistency between goals and system specifications. For example, Ghabi et al. \cite{31} proposed a verification framework that supports uncertainty representation, enabling the automatic derivation of logical links between models and implementations, thereby enhancing developers' deep understanding of system consistency across abstraction levels.

\noindent \textbullet\ \textbf{Regulatory Compliance Support:} This category of studies aims to ensure that software applications comply with regulatory requirement; e.g., to rebuild missing traceability links between \textit{requirement} and \textit{regulation} in the healthcare, finance, or insurance domains \cite{1,2}. The various studies targeting requirement artifact differ in the granularity of the requirement being analyzed.

The diversity of research scenarios reflects the multidimensional value of software traceability throughout software lifecycle. Many core application scenarios (e.g., change impact analysis, bug localization, code navigation, and program comprehension) directly support developers' daily tasks, indicating a major trend in current research: leveraging traceability to enhance development efficiency.

\subsubsection{Guidelines}
While our statistical analysis indicates that the technical foundations for link recovery are becoming increasingly sophisticated (with high academic output focusing on metrics such as precision and recall), the stark contrast between high academic and low industrial adoption suggests a critical alignment gap. Many traceability links are currently established in isolation, optimizing for statistical performance without a clear definition of who will use them or how they generate value. Such blindly established links incur continuous maintenance costs while delivering negligible utility, contributing to the traceability decay often observed in practice.

\begin{figure*}
    \centering
    \includegraphics[width=0.8\linewidth]{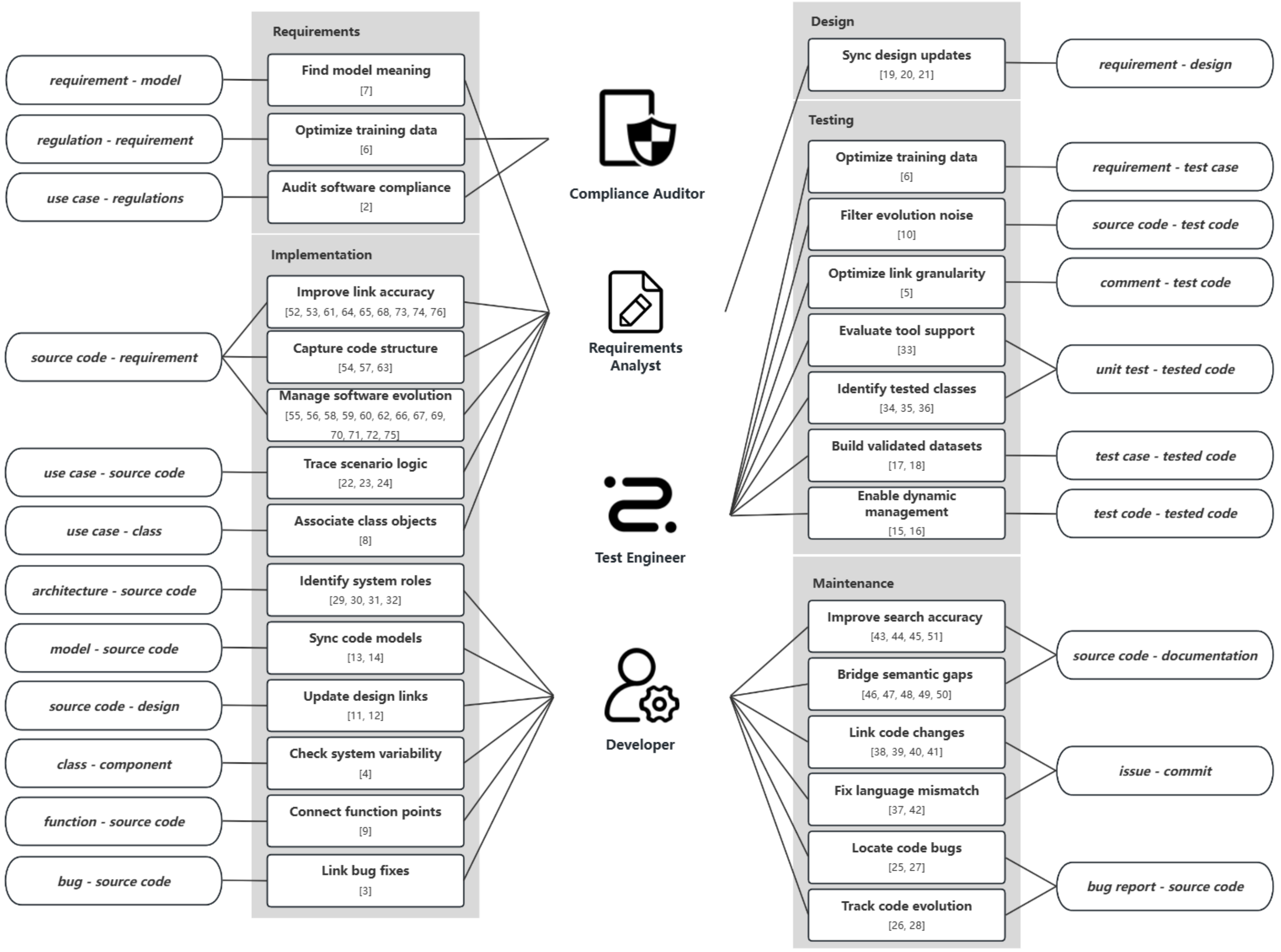}
    \caption{Goal-Driven Traceability Framework. This framework aligns specific traceability links with the lifecycle stages and core issues relevant to four key roles: auditors, analysts, testers, and developers.}
    \label{fig:role}
\end{figure*}

To address this, we propose a paradigm shift from ``artifact-centric'' to ``role-centric'' traceability, as illustrated in Fig. \ref{fig:role}. 
Specifically, we synthesized research objectives from the literature and mapped them to lifecycle stages and core issues relevant to four key roles: requirements analysts, developers, test engineers, and compliance auditors. This framework identifies the critical artifact associations that each role must prioritize to address their specific engineering challenges.

As visualized in the figure, traceability should not be viewed as a static graph of artifact pairs, but as a dynamic service layer supporting specific stakeholder needs. For example, for compliance auditors, links must prioritize auditability and provenance (e.g., ``\textit{regulation} - \textit{requirement}'') to minimize legal risk; for developers, links should focus on impact analysis and comprehension (e.g., ``\textit{source code} – \textit{design}'') to facilitate code evolution; and for test engineers, links must ensure coverage (e.g., ``\textit{requirement} – \textit{test case}'') to optimize regression testing.

Future research and practice should adhere to the principle of defining the goal before the link. Practitioners are advised to periodically review existing trace links to prune low-value associations, focusing resources strictly on the high-value paths identified in our goal-driven traceability framework.



\begin{mdframed}[style=graystyle]
\textbf{Answer to RQ3:} {Software traceability recovery studies were conducted mainly in an academic setting and covered the entire software lifecycle. The usage scenarios for recovered links are overwhelmingly concentrated on requirement-implementation consistency and maintenance support, which account for the majority of the studied literature.}
\end{mdframed}
\begin{table*}[]
\caption{Summary of Domains, Software Lifecycle Phases, and Objectives for Associations Between Artifacts}
\resizebox{\textwidth}{!}{
\begin{tabular}{|l|rr|c|l|}
\hline
\multicolumn{1}{|c|}{\multirow{2}{*}{\textbf{Software Artifacts A - Software Artifacts B}}} & \multicolumn{2}{c|}{\textbf{Domain}} & \multirow{2}{*}{\textbf{Software Lifecycle Phase}} & \multicolumn{1}{c|}{\multirow{2}{*}{\textbf{Objective}}} \\ \cline{2-3}
\multicolumn{1}{|c|}{} & \multicolumn{1}{c|}{\textbf{Academic}} & \multicolumn{1}{c|}{\textbf{Industrial}} & & \multicolumn{1}{c|}{} \\ \hline

source code - requirement (25) & \multicolumn{1}{r|}{25/25} & 0/25 & \multirow{5}{*}{\begin{tabular}[c]{@{}c@{}}Requirement\\ Design\\ Implementation\end{tabular}} & \multirow{5}{*}{\begin{tabular}[c]{@{}l@{}}\textbf{Requirements-Implementation Consistency (34) }mainly includes:\\ · Enhancing Accuracy and Quality of Automated Recovery (10)\\ · Bridging Semantic Gaps and Vocabulary Mismatches (9)\\ · Mitigating Traceability Degradation during Software Evolution (6)\end{tabular}} \\ \cline{1-3}
use case - source code (3) & \multicolumn{1}{r|}{3/3} & 0/3 & & \\ \cline{1-3}
requirement - design (3) & \multicolumn{1}{r|}{2/3} & 1/3 & & \\ \cline{1-3}
source code - design (2) & \multicolumn{1}{r|}{2/2} & 0/2 & & \\ \cline{1-3}
use case - class (1) & \multicolumn{1}{r|}{1/1} & 0/1 & & \\ \hline
issue - commit (6) & \multicolumn{1}{r|}{6/6} & 0/6 & \multirow{8}{*}{\begin{tabular}[c]{@{}c@{}}Requirement\\ Implementation\\ Testing\\ Maintenance\end{tabular}} & \multirow{8}{*}{\begin{tabular}[c]{@{}l@{}}\textbf{Quality Assurance and Maintenance (21)} mainly includes:\\ · Enhancing Recovery Accuracy and Semantic Alignment (7)\\ · Addressing Data Sparsity and Industrial Practicality (5)\end{tabular}} \\ \cline{1-3}
bug report - source code (4) & \multicolumn{1}{r|}{4/4} & 0/4 & & \\ \cline{1-3}
unit test - tested code (4) & \multicolumn{1}{r|}{4/4} & 0/4 & & \\ \cline{1-3}
test code - tested code (2) & \multicolumn{1}{r|}{2/2} & 0/2 & & \\ \cline{1-3}
test case - tested code (2) & \multicolumn{1}{r|}{2/2} & 0/2 & & \\ \cline{1-3}
bug - source code (1) & \multicolumn{1}{r|}{1/1} & 0/1 & & \\ \cline{1-3}
requirement - test case (1) & \multicolumn{1}{r|}{1/1} & 0/1 & & \\ \cline{1-3}
source code - test code (1) & \multicolumn{1}{r|}{1/1} & 0/1 & & \\ \hline
source code - documentation (9) & \multicolumn{1}{r|}{8/9} & 1/9 & \multirow{4}{*}{\begin{tabular}[c]{@{}c@{}}Design\\ Implementation\\ Testing\\ Maintenance\end{tabular}} & \multirow{4}{*}{\begin{tabular}[c]{@{}l@{}}\textbf{Program Comprehension (12)} mainly includes:\\ · Establishing Traceability for Informal and Legacy Environments (5)\end{tabular}} \\ \cline{1-3}
class - component (1) & \multicolumn{1}{r|}{1/1} & 0/1 & & \\ \cline{1-3}
function - source code (1) & \multicolumn{1}{r|}{1/1} & 0/1 & & \\ \cline{1-3}
comment - test code (1) & \multicolumn{1}{r|}{1/1} & 0/1 & & \\ \hline
architecture - source code (4) & \multicolumn{1}{r|}{4/4} & 0/4 & \multirow{3}{*}{\begin{tabular}[c]{@{}c@{}}Requirement\\ Design\\ Implementation\end{tabular}} & \multirow{3}{*}{\begin{tabular}[c]{@{}l@{}}\textbf{Model-driven System Understanding (7)} mainly includes:\\ · Reconstructing Architectural Knowledge from Limited Information (3)\end{tabular}} \\ \cline{1-3}
model - source code (2) & \multicolumn{1}{r|}{1/2} & 1/2 & & \\ \cline{1-3}
requirement - model (1) & \multicolumn{1}{r|}{0/1} & 1/1 & & \\ \hline
regulation - requirement (1) & \multicolumn{1}{r|}{1/1} & 0/1 & \multirow{2}{*}{Requirement} & \multirow{2}{*}{\begin{tabular}[c]{@{}l@{}}\textbf{Regulatory Compliance Support (2)} mainly includes:\\ · ensuring software application compliance to regulation (2)\end{tabular}} \\ \cline{1-3}
use case - regulation (1) & \multicolumn{1}{r|}{1/1} & 0/1 & & \\ \hline
\end{tabular}
}
\begin{flushleft}
{\footnotesize Numbers in parentheses indicate the total number of selected studies in each category. Values in the ``Academic'' and ``Industrial'' columns denote the number of papers in the respective domain out of the total papers.}
\end{flushleft}
\label{table:RQ3}
\end{table*}
\section{Discussion}\label{discussion}
Based on the analysis conducted in the previous sections, we distill several critical insights to provide a broader perspective on the current state of software traceability recovery. After that, we present the limitations of our study in this section.

\subsection{Takeaway}
Complementing the specific guidelines derived from RQ1 to RQ3, we highlight several takeaways for this community.

\textbf{From Static Recovery to Dynamic Evolution.} Early research treated traceability as a static and binary mapping task. However, our Artifacts Traceability Graph reveals a multidimensional ecosystem, demonstrating that isolated point-to-point recovery is structurally insufficient for evolving software. To prevent severe traceability decay, future research must transition from one-time mapping to continuous traceability maintenance, integrating automated recovery directly into CI/CD pipelines to ensure global consistency without incurring traceability debt.

\textbf{Breaking the Semantic Bottleneck.} Our analysis exposes a severe asymmetric maturity: traceability recovery excels in textual artifacts but largely neglects structural models and abstractions. This disparity creates a semantic bottleneck that leaves developers blind to whether specific code changes violate high-level architectural constraints. To prevent silent architectural drift, future research must move beyond traditional textual similarity. We urge the community to leverage cross-modal alignment and graph representation learning to project heterogeneous artifacts into a unified topological space.

\textbf{Balancing Semantic Power, Reliability, and Efficiency.} Our RQ2 analysis reveals that traceability techniques are rapidly evolving from cost-efficient lexical matching to advanced semantic models (e.g., PLMs/LLMs). However, this shift introduces critical challenges: massive computational overhead and the severe risk of \textit{hallucinations} (i.e., generating plausible but incorrect links). To achieve trustworthy and sustainable traceability, future research can develop hybrid approaches that constrain large models' outputs with deterministic rules, or compress the deep contextual reasoning of large models into lightweight and locally-deployable tools for daily engineering workflows (e.g., through knowledge distillation).

\subsection{Threats to Validity}
We discuss the threats to the validity of this review across three dimensions.

\textbf{Construct Validity.} The diverse and evolving terminology in software traceability introduces a risk of missing relevant literature during keyword-based searches. To mitigate this threat, we systematically queried five major databases and supplemented our retrieval with a bidirectional snowballing strategy to capture elusive studies.

\textbf{Internal Validity.} The manual literature filtering and artifact classification inevitably introduce subjective selection bias. We minimized this by establishing rigorous inclusion criteria and employing independent cross-validation by two authors, with a third resolving conflicts. Furthermore, our linking tool analysis relies entirely on self-reported data from the primary studies rather than independent tool execution.

\textbf{External Validity.} The generalizability of our proposed frameworks may be limited. Although validated by an expert survey, the results reflect participants' perceived usefulness rather than actual performance in complex industrial settings. Future large-scale controlled experiments are required to validate their practical efficacy.





\section{Conclusions}\label{conclusion}
In this paper, we conducted a systematic literature review to provide a comprehensive overview of software traceability recovery. We collected 76 research papers that recovered traceability links between software artifacts. We summarized 22 types of software artifacts and 23 types of associations analyzed in the current research.
We confirmed that diverse artifacts, predominantly \textit{source code}, \textit{test code}, and \textit{requirement}, are focal points for traceability. Additionally, our analysis highlighted a significant shift towards employing advanced learning models for their advanced semantic understanding, complementing traditional techniques in establishing these crucial links. However, the current evaluation revealed critical limitations: the low availability of source code (only 37\% of studies) and the absence of standardized evaluation metrics. We also found that the usage scenarios for recovered links are overwhelmingly concentrated on requirement-implementation consistency and maintenance support (72\% of studies).

Future research should prioritize developing robust tools with openly shared source code, establishing unified metrics for performance evaluation, and exploring the full potential of advanced AI techniques, to address the remaining challenges in global and goal-driven traceability recovery scenarios. 

\bibliographystyle{IEEEtran}
\bibliography{txt/acmart.bib}

\end{document}